\documentclass[12pt]{article}
\usepackage{amsfonts}
\usepackage{amsmath}
\usepackage{amssymb}
\usepackage{mathptmx}
\usepackage[dvips]{color}
\usepackage{epsfig}
\usepackage{graphicx}
\newcommand{\calA}{{\cal A}}

\newcommand{\calL}{{\cal L}}

\newcommand{\calO}{{\cal O}}

\newcommand{\dslash}{/\!\!\!\!\!\partial}

\newcommand{\pslash}{/\!\!\!\!\!p}

\newcommand{\GeV}{{\rm GeV}}

\newcommand{\cm}{{\rm cm}}
\newcommand{\kpc}{{\rm kpc}}
\newcommand{\pc}{{\rm pc}}

\begin{document}
\baselineskip=16pt

\pagenumbering{arabic}

\vspace{1.0cm}

\begin{center}
{\Large\sf Spin 3/2 particle as a dark matter candidate: an
effective field theory approach}
\\[10pt]
\vspace{.5 cm}

{Ran Ding $^a$\footnote{dingran@mail.nankai.edu.cn} and Yi Liao
$^{a,b,c}$\footnote{liaoy@nankai.edu.cn}}

{\it $^a$ School of Physics, Nankai University, Tianjin 300071,
China
\\
$^b$ Center for High Energy Physics, Peking University, Beijing
100871, China
\\
$^c$ Kavli Institute for Theoretical Physics China, CAS, Beijing
100190, China}

\vspace{2.0ex}

{\bf Abstract}

\end{center}

There is no indication so far on the spin of dark matter particles.
We consider the possibility in this work that a spin-3/2 particle
acts as dark matter. Employing the approach of effective field
theory, we list all possible 4-fermion effective interactions
between a pair of such fields and a pair of ordinary fermion fields.
We investigate the implications of the proposal on the relic
density, the antiproton to proton flux ratio in cosmic rays, and the
elastic scattering off nuclei in direct detection. While the relic
density and flux ratio are sensitive to all interactions albeit at
different levels, the direct detection is only sensitive to a few of
them. Using the observed data and experimental bounds, we set
constraints on the relation of couplings and dark particle mass. In
particular, we find that some mass ranges can already be excluded by
jointly applying the observed relic density on the one side and the
measured antiproton to proton flux ratio or the upper bounds from
direct detection on the other.

%
%
%



\newpage

\section{Introduction}

There is compelling evidence from astronomical observations that the
dominant component of matter in our universe is invisible, dubbed
dark matter (DM). After years of efforts the evidence is still
mainly restricted to the scope of gravitational effects. There are
now many on-going or approved astronomical and laboratory projects
that will hopefully reveal in the near future whether the dark
matter is composed of particles or astrophysical objects like
massive compact halo objects or it is not required at all because of
modified Newton dynamics. For brief overviews on the current
experimental status, see for instance the recent talks in Ref.
\cite{Strigari:2010zz}.

From the theoretical point of view we are not short of candidates if
the dark matter turns out to be composed of particles. The most
popular one is the lightest supersymmetric particle, perhaps the
lightest neutralino, in supersymmetric models. There are also
extensive discussions suggesting that the lightest Kaluza-Klein
particle in extra dimension models \cite{Kolb:1983fm} or the
lightest T-odd particle in little Higgs models with T parity
\cite{Cheng:2004yc} could act as DM, and so on. For detailed
reviews, see for instance, Ref. \cite{Jungman:1995df}. Since the
models on which the proposals mentioned are based are yet to be
verified, it is important not to forget about other alternatives. In
this context, effective field theory serves as a useful approach
since one can focus on the interactions relevant to DM searches and
parameterize unknown underlying dynamics in terms of effective
couplings \cite{Goodman:1984dc}, while leaving the dynamics to be
identified in dedicated particle experiments like high energy
colliders.

The effective field theory approach has been widely employed to
study the detection of a scalar, spin-1/2 fermion and vector DM
particle
\cite{Kurylov:2003ra,Birkedal:2004xn,Giuliani:2004uk,Feng:2005gj,Belanger:2008sj,
Beltran:2008xg,Cirelli:2008pk,Cao:2009uw,Agrawal:2010fh,Fan:2010gt,
Goodman:2010ku,Cheung:2010zf,Cheung:2010ua,Zheng:2010js,Frandsen:2011cg}.
In this work we consider the possibility of a spin-3/2 particle as
a DM candidate, and investigate its features in various detection
experiments. Although we do not know yet a quantum field theory
for such a particle of a higher spin that would be renormalizable
in the conventional sense, this is by no means a reason to exclude
its physical relevance: there are hadronic resonances of higher
spin, e.g., $\Delta(1232)$ of spin $3/2$, which play an important
role in nuclear physics. Another example of spin-3/2 particles is
the well-motivated hypothetical gravitino, which is a gauge
particle associated with spontaneously broken localized
supersymmetry. We are aware that even the effective interactions
involving the $\Delta(1232)$ resonance are still controversial,
see for instance \cite{Haberzettl:1998rw} for a summary of the
relevant issues. Our results in this work are nevertheless immune
from such uncertainties since what we will need is not more than
the Lorentz covariance from which the polarization sum of a
spin-3/2 particle is constructed. We also know that the gravitino
itself was previously suggested as a DM particle. It interacts
with ordinary particles with an essentially gravitational
strength, and the leading interactions at low energies are those
involving a single gravitino field (and another superpartner); see
Ref. \cite{Moroi:1995fs} for a practical introduction to the
relevant issues. In contrast, what we will consider in the
following are the effective interactions that contain a pair of
spin-3/2 fields. Such interactions naturally preserve certain
parity if the latter is required to stabilize the DM particle.

The paper is organized as follows. In the next section we exhaust
all possible 4-fermion effective interactions that involve a pair of
spin-3/2 fields and a pair of ordinary spin-1/2 fields. We then
compute the cross sections for the annihilation and elastic
scattering processes. These results will be utilized in section 3 to
compute the secondary particle fluxes in the cosmic rays, the relic
density, and the effective cross sections between the DM particles
and nucleons. Using the observational and laboratory data we set
constraints on the couplings of the effective interactions as a
function of the DM mass. It turns out some of the mass ranges can be
excluded using the currently available data. We recapitulate our
results in the last section.

\section{Effective interactions and cross sections}

Suppose the stability of the spin-3/2 particle is protected by
certain parity, its leading effective interactions would involve a
pair of it. Amongst the possible interactions that are relevant to
their detection are those that couple to a pair of ordinary
fermions. We therefore restrict ourselves in this work to the
4-fermion interactions.

A free particle of spin $3/2$ and mass $M$ can be described by a
field $\Psi_\mu$ that has the mixed transformation properties of a
Dirac four-component field and of a four-component vector field
\cite{Rarita:1941mf}. Its equation of motion is
\begin{eqnarray}
(i\dslash-M)\Psi_\mu=0,%
\label{eq_EoM}
\end{eqnarray}
augmented with the constraint, $\gamma^\mu\Psi_\mu=0$. Multiplying
eq (\ref{eq_EoM}) from the left by $\gamma^\mu$ and applying the
constraint one gets $\partial^\mu\Psi_\mu=0$ as a consequence. The
wavefunction of such a particle satisfies the same equations and
thus has four independent components as desired, that correspond
to the four spin states in its rest frame. As a matter of fact,
the wavefunction for a particle with four-momentum $p$ and
helicity $\lambda$, $U_\mu(p,\lambda)$, can be constructed using
the Clebsch-Gordan coefficients in terms of the ones for a
spin-1/2 Dirac spinor $u(p,s)$ of helicity $s$ and a spin-1
polarization $\epsilon(p,\kappa)$ of helicity $\kappa$
\cite{Kusaka:1941}. Our following calculation will not depend on
the explicit form of $U_\mu$ but its polarization sum
$P_{\nu\mu}(p)=\sum_\lambda U_\nu(p,\lambda)\bar U_\mu(p,\lambda)$
which is evaluated to be (see for instance, Ref.
\cite{Moroi:1995fs} upon correcting the sign of the $M$ term)
\begin{eqnarray}
P_{\mu\nu}(p)=-(\pslash+M)\bigg(T_{\mu\nu}(p)-
\frac{1}{3}\gamma^\rho
T_{\rho\mu}(p)T_{\nu\sigma}(p)\gamma^\sigma\bigg),%
\label{eq_pol_sum}
\end{eqnarray}
with $T_{\mu\nu}(p)=g_{\mu\nu}-p_\mu p_\nu/p^2$ and $p^2=M^2$. Note
that the factor $(\pslash+M)$ can be equally well put on the
rightmost. The equation of motion and the constraint imply that
\begin{eqnarray}
&&\gamma^\mu P_{\mu\nu}(p)=P_{\mu\nu}(p)\gamma^\nu=0,
\nonumber\\
&&p^\mu P_{\mu\nu}(p)=P_{\mu\nu}(p)p^\nu=0,
\nonumber\\
&&(\pslash-M)P_{\mu\nu}(p)=P_{\mu\nu}(p)(\pslash-M)=0,
\end{eqnarray}
which may be employed to verify eq (\ref{eq_pol_sum}). The
polarization sum $Q_{\mu\nu}(p)$ for the antiparticle and its
relations can simply be obtained from the above by $M\to-M$. We
will assume in this work that the spin-3/2 particle is of Dirac
nature. For a Majorana-type particle, the amplitudes to be
computed later are either multiplied by a factor of two or just
vanish.

The leading 4-fermion interactions between a pair of $\Psi_\mu$
fields and a pair of ordinary fermion fields $f$ are of dimension
six. For simplicity we will not consider flavor-changing
interactions. The independent Hermitian bilinears of $f$ are
\begin{eqnarray*}
(a):~\bar ff,~\bar fi\gamma_5f,~\bar f\gamma_\mu f,~\bar
f\gamma_\mu\gamma_5f,~\bar f\sigma_{\mu\nu}f.
\end{eqnarray*}
The bilinears of $\Psi_\mu$ are similar with the only difference in
that they have two additional vector indices:
\begin{eqnarray*}
(i):~\bar\Psi_\alpha\Psi_\beta,~\bar\Psi_\alpha i\gamma_5\Psi_\beta,
~\bar\Psi_\alpha\gamma_\rho\Psi_\beta,
~\bar\Psi_\alpha\gamma_\rho\gamma_5\Psi_\beta,
~\bar\Psi_\alpha\sigma_{\rho\sigma}\Psi_\beta.
\end{eqnarray*}
Consider the self-contraction of a pair of indices in the list
$(i)$. Since the interactions will be exploited in the cases where
the spin-3/2 particles are on-shell, the constraint for the free
field still applies. This means that it is not necessary to
consider the contraction between the fields and the $\gamma$
matrices. For instance, after a little algebra, we find
$g^{\alpha\rho}\bar\Psi_\alpha\sigma_{\rho\sigma}\Psi_\beta
=-i\bar\Psi_\sigma\Psi_\beta$, which however was already covered
in the list $(i)$. The contraction is thus restricted to be
between the two factors of the $\Psi_\mu$ field:
\begin{eqnarray*}
(ii):~\bar\Psi_\alpha\Psi^\alpha,~\bar\Psi_\alpha
i\gamma_5\Psi^\alpha, ~\bar\Psi_\alpha\gamma_\rho\Psi^\alpha,
~\bar\Psi_\alpha\gamma_\rho\gamma_5\Psi^\alpha,
~\bar\Psi_\alpha\sigma_{\rho\sigma}\Psi^\alpha,
\end{eqnarray*}
while further contraction produces nothing new.

All possible interactions are exhausted by multiplying the terms in
the list $(a)$ with those in the list $(i)$ and $(ii)$ respectively
and contracting remaining indices with the signature tensor
$g^{\mu\nu}$ or the totally antisymmetric tensor
$\epsilon^{\mu\nu\rho\sigma}$. Some of the terms so obtained can be
removed as redundant. For instance, using
$\epsilon^{\mu\nu\rho\sigma}\sigma_{\rho\sigma}=-i2\sigma^{\mu\nu}\gamma_5$
(in our convention $\gamma_5=i\gamma^0\gamma^1\gamma^2\gamma^3$ and
$\epsilon^{0123}=+1$), we have $\epsilon^{\alpha\beta\rho\sigma}\bar
ff\bar\Psi_\alpha\sigma_{\rho\sigma}\Psi_\beta =-2\bar
ff\bar\Psi_\alpha\gamma_5\Psi^\alpha$,
$\epsilon^{\alpha\beta\rho\sigma}\bar
fi\gamma_5f\bar\Psi_\alpha\sigma_{\rho\sigma}\Psi_\beta =-2\bar
fi\gamma_5f\bar\Psi_\alpha\gamma_5\Psi^\alpha$, and so on. The final
list contains the following 14 operators:
\begin{eqnarray}
\calO^f_{1,\dots,4}&=&\bar f\sigma_{\mu\nu}f\big(\bar\Psi^\mu i\Psi^\nu,~%
\bar\Psi^\mu \gamma_5\Psi^\nu\big),~%
\bar f\sigma^{\mu\nu}i\gamma_5f\big(\bar\Psi_\mu i\Psi_\nu,
~\bar\Psi_\mu\gamma_5\Psi_\nu\big);
\nonumber\\
\calO^f_{5,\dots,8}&=&\bar ff\big(\bar\Psi_\alpha\Psi^\alpha,~
\bar\Psi_\alpha i\gamma_5\Psi^\alpha\big),~%
\bar fi\gamma_5f\big(\bar\Psi_\alpha\Psi^\alpha,~%
\bar\Psi_\alpha i\gamma_5\Psi^\alpha\big);
\nonumber\\
\calO^f_{9,\dots,12}&=&%
\bar f\gamma_\mu f\big(\bar\Psi_\alpha\gamma^\mu\Psi^\alpha,~%
\bar\Psi_\alpha\gamma^\mu\gamma_5\Psi^\alpha\big),~%
\bar f\gamma_\mu\gamma_5 f\big(\bar\Psi_\alpha\gamma^\mu\Psi^\alpha,~%
\bar\Psi_\alpha\gamma^\mu\gamma_5\Psi^\alpha\big);
\nonumber\\
\calO^f_{13,14}&=&\big(\bar f\sigma_{\mu\nu}f,~%
\bar f\sigma_{\mu\nu}i\gamma_5f\big)\bar
\Psi_\alpha\sigma^{\mu\nu}\Psi^\alpha.
\end{eqnarray}
The effective interactions are summarized as
\begin{eqnarray}
\calL_\textrm{int}&=&\sum_f\sum_{i=1}^{14}G_i^f\calO^f_i,
\label{eq_L}
\end{eqnarray}
where all couplings $G_i^f$ are real and have the same dimensions as
the Fermi constant $G_F$.

To prepare for the numerical analysis in the next section, we
display the cross sections for DM annihilation and elastic
scattering off a nucleus. Since $\calL_\textrm{int}$ contains many
possible interactions, it looks sensible to treat one interaction at
a time. When $\Psi$ in a specific model happens to interact in
several ways with an ordinary fermion of a given flavor, one should
sum coherently their contributions to a scattering amplitude. The
spin-summed and -averaged cross section in the center-of-mass frame
for the annihilation process $\Psi\bar\Psi\to f\bar f$ through the
interaction $\calO_i^f$, is
\begin{eqnarray}
\sigma_i^f&=&N^f\frac{\big(G_i^f\big)^2s}{16\pi}
\sqrt{\frac{s-4m_f^2}{s-4M^2}}A_i(m_f^2/s,M^2/s),%
\label{eq_anni}
\end{eqnarray}
where $s$ is the center-of-mass energy squared, $m_f$ and $M$ are
respectively the masses of the final ($f$) and initial ($\Psi$)
particles, and $N^f=1$ (3) when $f$ is a lepton (quark). The
dimensionless functions $A_i$ for various operators are listed in
the Appendix.

In direct detection of dark matter one measures the recoil energy of
nuclei that have been struck by a DM particle in the local halo. The
event rate and energy deposited are determined by the cross section
between the two. The calculation of the latter is a hard task,
connecting microscopic interactions of DM particles with quarks to
effective interactions with nuclei through the intermediate chiral
dynamics of nucleons, incurring uncertainties at each step; see the
first article in Ref. \cite{Jungman:1995df} for a review. The
difficulty is alleviated to some extent by the fact that the
collision is nonrelativistic. In this case the DM particles only
feel the spin or mass of a nucleus \cite{Goodman:1984dc}. In the
remainder of this section we outline the procedure of this
calculation relevant to our case and present the results for cross
sections between $\Psi$ and a nucleus.

To begin with, one builds effective interactions between $\Psi$ and
nucleons from those for quarks shown in eq (\ref{eq_L}). One assumes
that the Lorentz structures are not changed but the interaction
strengths get corrected by chiral dynamics. Since the nucleons in a
nucleus can be treated nonrelativistic for the purpose here, the
relevant nucleon bilinears (and thus quark bilinears) are restricted
to the spin-independent (SI) and spin-dependent (SD) ones. In the
approximation of zero momentum transfer, one then takes the diagonal
matrix element of the nucleon bilinears for a static nucleus. The SI
part essentially counts the numbers of the protons and neutrons in
the nucleus while the SD part gets its main contribution from the
unpaired nucleon spin. For a reasonable estimate of the nuclear
matrix elements one has to appeal to nuclear models; in particular,
when the momentum transfer must be taken into account, a form factor
is necessary that will reduce the cross section with nuclei.

Following the above procedure, the effective interactions in eq
(\ref{eq_L}) are first classified into the SI and SD parts for
nonrelativistic quarks:
\begin{eqnarray}
\calL^\textrm{SI}_{x,\textrm{s}}&=&B_x(\Psi)\sum_qG^q_xS_q,~x=5,6,
\\
\calL^\textrm{SI}_{y,\textrm{v}}&=&B_y(\Psi)\sum_qG^q_yV_q,~y=9,10,
\\
\calL^\textrm{SD}_z&=&B^k_z(\Psi)\sum_qG^q_zA_q^k,~z=1,\dots,4,11,\dots,14,
\end{eqnarray}
where $S_q=\bar qq$, $V_q=\bar q\gamma_0q$, $A_q^k=\bar
q\gamma^k\gamma_5q$, and $B_{x,y}(\Psi)$ and $B_z^k(\Psi)$ are the
$\Psi$ bilinears without or with a free spatial index
respectively. Although the nucleon matrix elements of the
pseudoscalar quark bilinears do not vanish (see the second paper
in Ref. \cite{Cheng:1988cz}), the induced pseudoscalar nucleon
bilinear has a nuclear matrix element that is suppressed by the
velocity of nucleons. Thus the operators $\calO_{7,8}$ do not
contribute at leading order to the scattering of $\Psi$ off a
nucleus $N$ as in the usual practice. The scattering amplitudes
for the $\Psi$-$N$ scattering are
\begin{eqnarray}
\calA^\textrm{SI}_{x,\textrm{s}}&\approx&2m_Nf_N^xB_x(U),
\\
\calA^\textrm{SI}_{y,\textrm{v}}&\approx&2m_Nb_N^yB_y(U),
\\
\calA^\textrm{SD}_z&\approx&4m_Ng_N^z(J_N^k)_{fi}B^k_z(U),
\end{eqnarray}
where $m_N$ is the mass of the nucleus with atomic mass number $A$
and charge $Z$, and $(J_N^k)_{fi}$ is the matrix element of the
$k$-th component of the nuclear spin operator. The scalar SI
effective coupling gets contributions from the protons and neutrons
contained in $N$,
\begin{eqnarray}
f_N^x&=&Zf_p^x+(A-Z)f_n^x,
\\
f_{p(n)}^x&=&\sum_{q=u,d,s}G^q_x\frac{m_{p(n)}}{m_q}f_{Tq}^{p(n)}
+\frac{2}{27}f_{TG}^{p(n)}\sum_{q=c,b,t}G^q_x\frac{m_{p(n)}}{m_q},
\end{eqnarray}
where $m_{p(n,q)}$ is the proton (neutron, quark) mass. For light
quarks the constants $f_{Tq}^{p(n)}$ are related to the pion-nucleon
sigma term \cite{Cheng:1988cz}, while for heavy quarks
$\displaystyle f_{TG}^{p(n)}=1-\sum_{q=u,d,s}f^{p(n)}_{Tq}$ enter
via the trace anomaly \cite{Shifman:1978zn}. The vector SI effective
coupling is easiest to get since $V_q$ just counts the number of
valence quarks when sandwiched between the nucleon states:
\begin{eqnarray}
b_N^y=Zb_p^y+(A-Z)b_n^y,~b_p^y=2G_y^u+G_y^d,~b_n^y=G_y^u+2G_y^d.
\end{eqnarray}
Finally, the SD effective coupling is,
\begin{eqnarray}
g_N^z&=&\sum_qG^q_z\lambda_q^N,
\\
\lambda_q^N&=&\frac{1}{J_N}\big[\langle
S_p\rangle\Delta_q^p+\langle S_n\rangle\Delta_q^n\big],
\end{eqnarray}
where $\Delta_q^{p(n)}$ measures the fraction of the proton
(neutron) spin carried by the quark $q$ \cite{QCDSF:2011aa}, and
$\langle S_{p(n)}\rangle$ is the expectation value of the $z$-th
component proton (neutron) spin operator in the nuclear state with
the highest $J_N^z$ \cite{Engel:1989ix}.

The spin-summed and -averaged cross sections for $\Psi$-$N$
scattering at zero momentum transfer are,
\begin{eqnarray}
\sigma_0=\frac{1}{16\pi(M+m_N)^2}\sum_\textrm{spins}
\overline{|\calA|^2},
\end{eqnarray}
where the amplitudes squared are evaluated in the standard manner
\begin{eqnarray}
\sum_\textrm{spins}\overline{|\calA^\textrm{SI}_{x,\textrm{s}}|^2}&=&
(2m_Nf_N^x)^2\frac{1}{4}\sum_{\Psi\textrm{ spins}}B_x(U)
B^\dagger_x(U),
\\
\sum_\textrm{spins}\overline{|\calA^\textrm{SI}_{y,\textrm{v}}|^2}&=&
(2m_Nb_N^y)^2\frac{1}{4}\sum_{\Psi\textrm{ spins}}B_y(U)
B^\dagger_y(U),
\\
\sum_\textrm{spins}\overline{|\calA^\textrm{SD}_z|^2}&=&
(4m_Ng_N^z)^2J_N(J_N+1)\frac{1}{4}\sum_{\Psi\textrm{
spins}}B^k_z(U)B^{k\dagger}_z(U).
\end{eqnarray}
The $\Psi$ spin sums are computed using its polarization sum. In
the nonrelativistic limit, this is facilitated by noting that only
the spatial components are nonvanishing
\begin{eqnarray}
P_{ij}(p)&=&M(\gamma_0+1)\Big(\delta_{ij}+\frac{1}{3}\gamma_i\gamma_j\Big).
\end{eqnarray}
The end results for the SI and SD cross sections due to various
interactions are respectively
\begin{eqnarray}
\sigma_0^5&=&\frac{\mu^2}{\pi}\big(f_N^5\big)^2,%
\label{eq_SIs}
\\
\sigma_0^9&=&\frac{\mu^2}{\pi}\big(b_N^9\big)^2,%
\label{eq_SIv}
\\
\sigma_0^{1,12,13}&=&\frac{\mu^2}{\pi}J_N(J_N+1)\big(g_N^{1,12,13}\big)^2
\times\left[\frac{20}{3},\frac{20}{3},\frac{80}{3}\right],%
\label{eq_SD}
\end{eqnarray}
where $\mu=m_NM/(m_N+M)$ is the reduced mass for the $\Psi$-$N$
system. That other operators do not contribute in the
nonrelativistic limit can also be understood explicitly. Since
$U_\mu(p,\lambda)$ is built from $u(p,s)$ and
$\epsilon_\mu(p,\kappa)$, its static limit can be readily
obtained. We have $U_0(p,\lambda)=0$ either from $p^\mu
U_\mu(p,\lambda)=0$ or by choosing physical polarizations with
$\epsilon_0=0$. Independently of the Lorentz index in $U_\mu$, the
limits for a Dirac spinor bilinear also apply to $U_\mu$, with the
nonvanishing bilinears being restricted to $\bar
U_\alpha\gamma^0U_\beta\approx\bar U_\alpha U_\beta$, $\bar
U_\alpha\gamma^i\gamma^5U_\beta$, $\bar
U_\alpha\sigma^{ij}U_\beta\approx\epsilon^{ijk} \bar
U_\alpha\gamma^k\gamma^5U_\beta$, $\bar
U_\alpha\sigma^{0i}\gamma^5U_\beta\approx i\bar
U_\alpha\gamma^i\gamma^5U_\beta.$ Thus $\calO^q_{2,4,6,7,8}$
vanish as they involve a pseudoscalar bilinear in $U_\mu$ or $q$,
while $\calO^q_3$ disappears since it couples a nonvanishing $q$
bilinear to $\Psi_0$. Similarly, $\calO^q_{10,11,14}$ do not
contribute either at the leading order.

\section{Constraints from observations and experiments}

\subsection{Relic density\label{sec-relic}}

The dark matter produced in the early universe would either be
depleted too much or over dense in the current epoch, depending on
the interaction strengths in eq (\ref{eq_L}). The observed value for
its relic density can therefore set constraints on the relevant
parameters. To obtain the relic number density $n_\Psi$, one solves
the Boltzmann equation
\begin{eqnarray}
\frac{dn_\Psi}{dt}+3Hn_\Psi=-\langle\sigma|v|\rangle
\big[(n_\Psi)^2-(n_\Psi^\textrm{eq})^2\big],%
\label{eq_Boltzmann}
\end{eqnarray}
where $H=\sqrt{8\pi\rho/3 M_\textrm{Pl}^2}$ is the Hubble expansion
rate and $n_\Psi^\textrm{eq}$ is the value at thermal equilibrium.
Assuming the DM particles have negligible chemical potential we have
$n_\Psi=n_{\bar\Psi}$ so that $n_\textrm{DM}=2n_\Psi$.
$\langle\sigma|v|\rangle$ is the thermally averaged annihilation
cross section for a relative velocity $v$, and can be calculated in
the reference frame where one of the $\Psi$ particles is at rest
\cite{Gondolo:1990dk}. Using
$s=2M^2\big[1+(1-v^2)^{-1/2}\big]\approx 4M^2(1+v^2/4)$ for
nonrelativistic DM particles, one expands eq (\ref{eq_anni}) as,
$\sigma|v|=a+bv^2+O(v^4)$. Eq (\ref{eq_Boltzmann}) is then solved
numerically to yield \cite{Kolbbook},
\begin{eqnarray}
\Omega_\textrm{DM}h^2\approx\frac{2\times 1.04\times
10^9x_F~\GeV^{-1}}{M_\textrm{Pl}\sqrt{g_*(x_F)}(a+3b/x_F)},
\end{eqnarray}
where $g_*(x_F)$ is the number of relativistic degrees of freedom at
the freeze-out temperature $T_F$, and $x_F=M/T_F$. The latter is
solved self-consistently by
\begin{eqnarray}
x_F=\ln\bigg[c(c+2)\sqrt{\frac{45}{8}}\frac{gMM_\textrm{Pl}(a+6b/x_F)}
{2\pi^3\sqrt{g_*(x_F)}}\bigg],
\end{eqnarray}
where $c$ is an order one parameter (we take $c=1/2$), and $g=4$ is
the spin degrees of freedom of the $\Psi$ particle.We employ the
values of $g_*$ as a function of temperature $T$ obtained in Ref.
\cite{Coleman:2003hs}.

\begin{figure}[!htbp]
\centering
\includegraphics[width=0.44\textwidth]{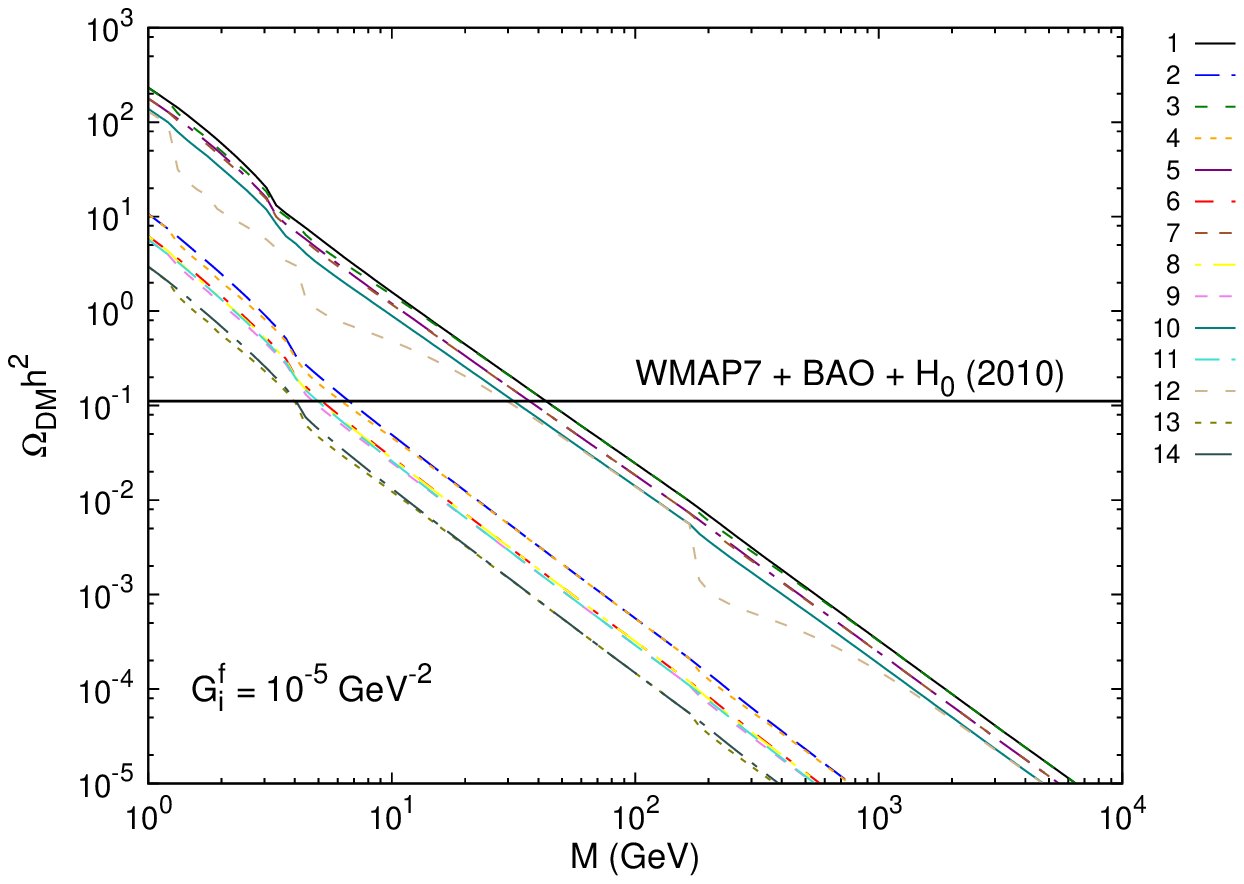}%
\hspace{0.01\textwidth}%
\includegraphics[width=0.44\textwidth]{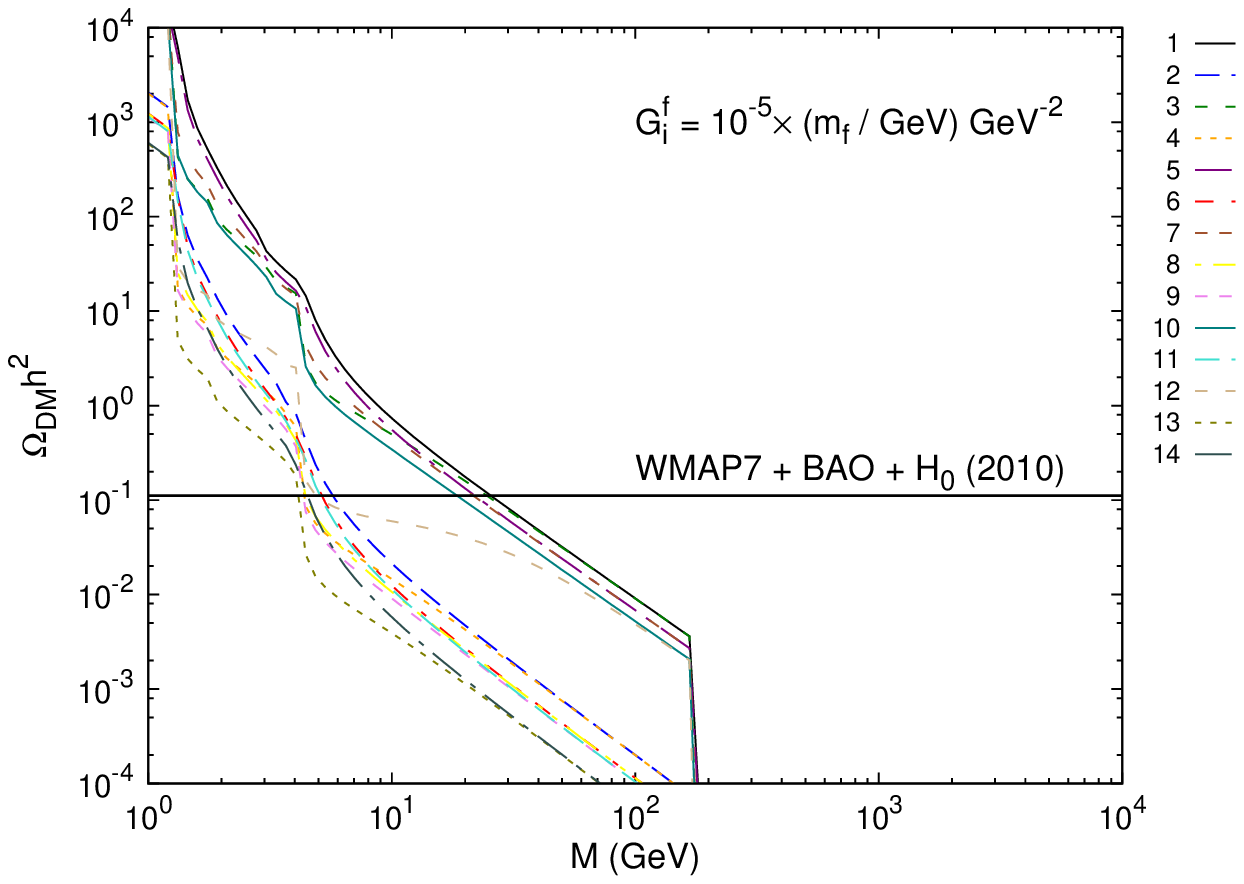}%
\\
\includegraphics[width=0.44\textwidth]{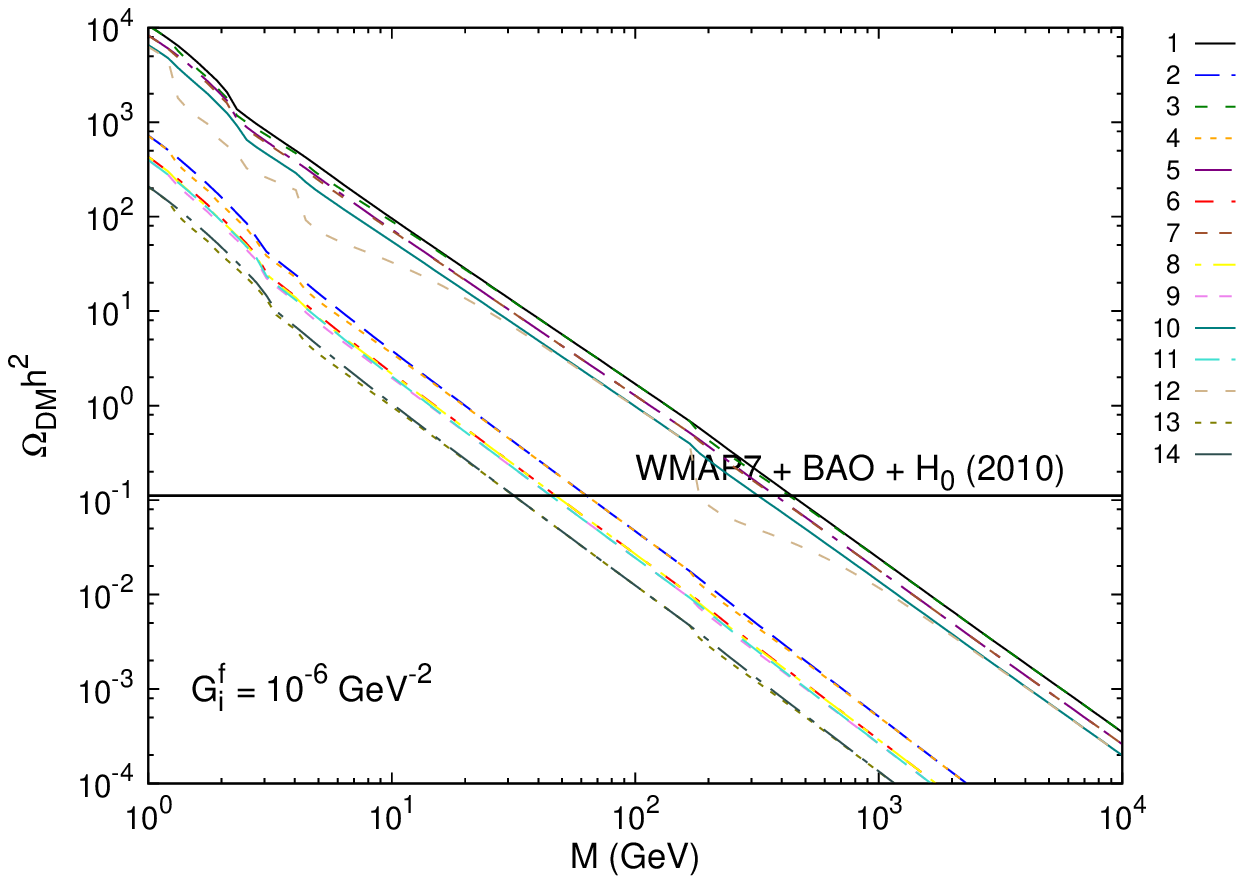}%
\hspace{0.01\textwidth}%
\includegraphics[width=0.44\textwidth]{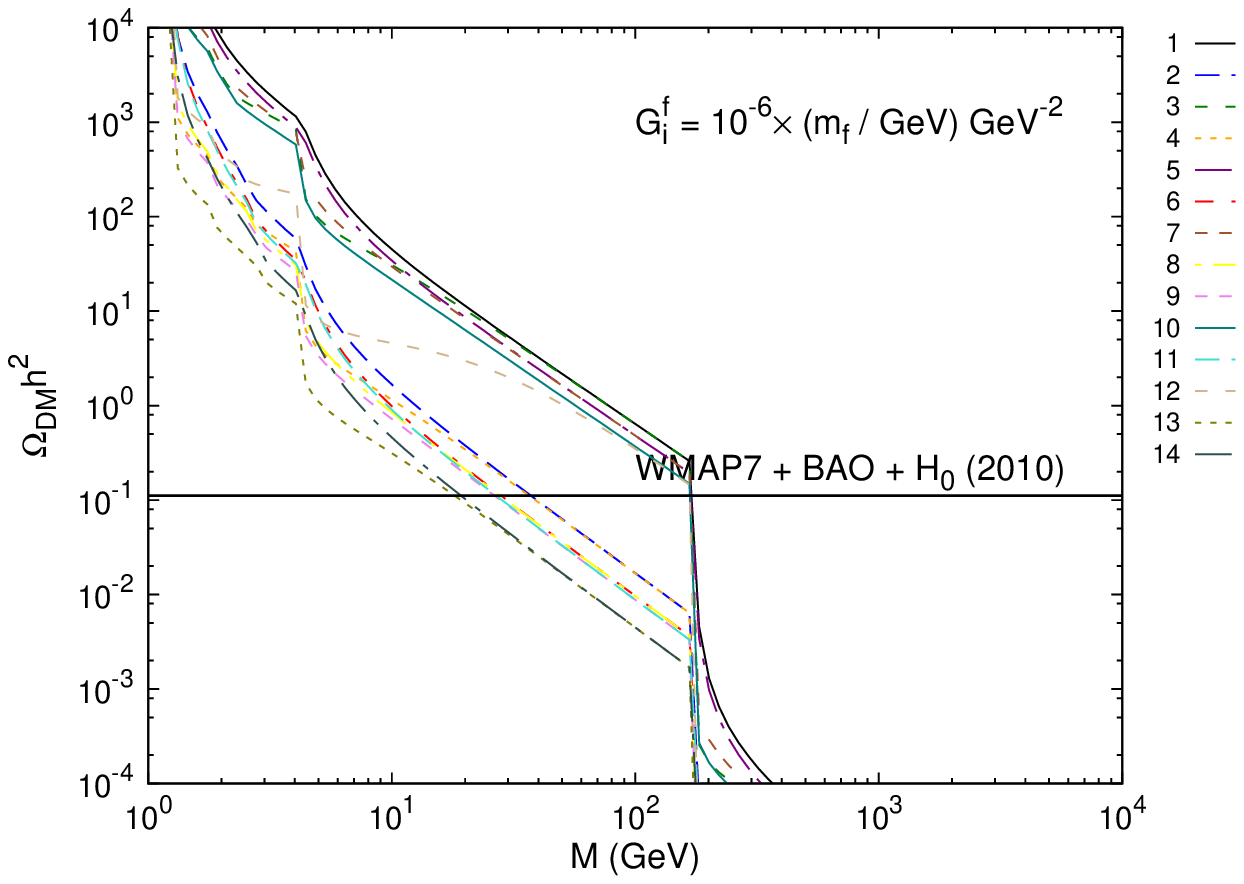}%
\\
\includegraphics[width=0.44\textwidth]{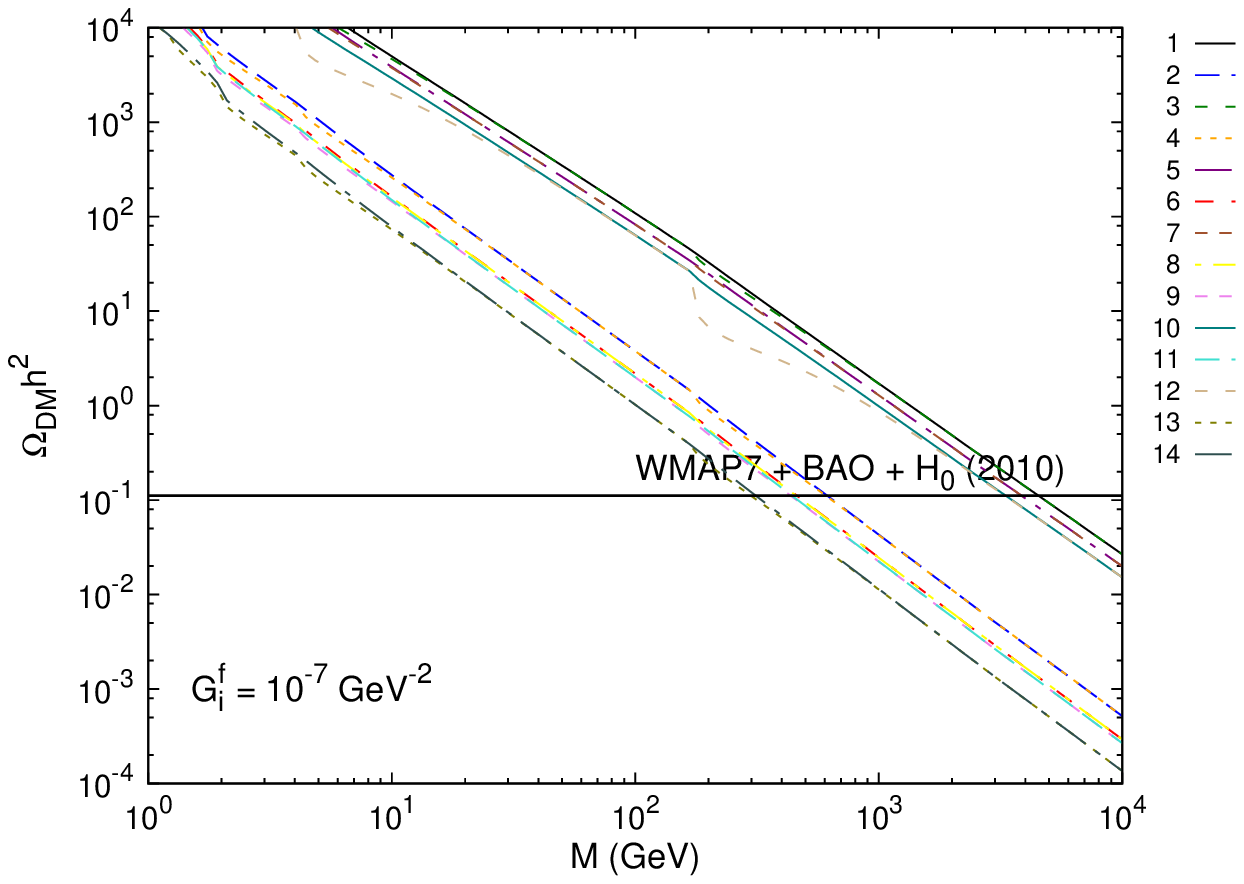}%
\hspace{0.01\textwidth}%
\includegraphics[width=0.44\textwidth]{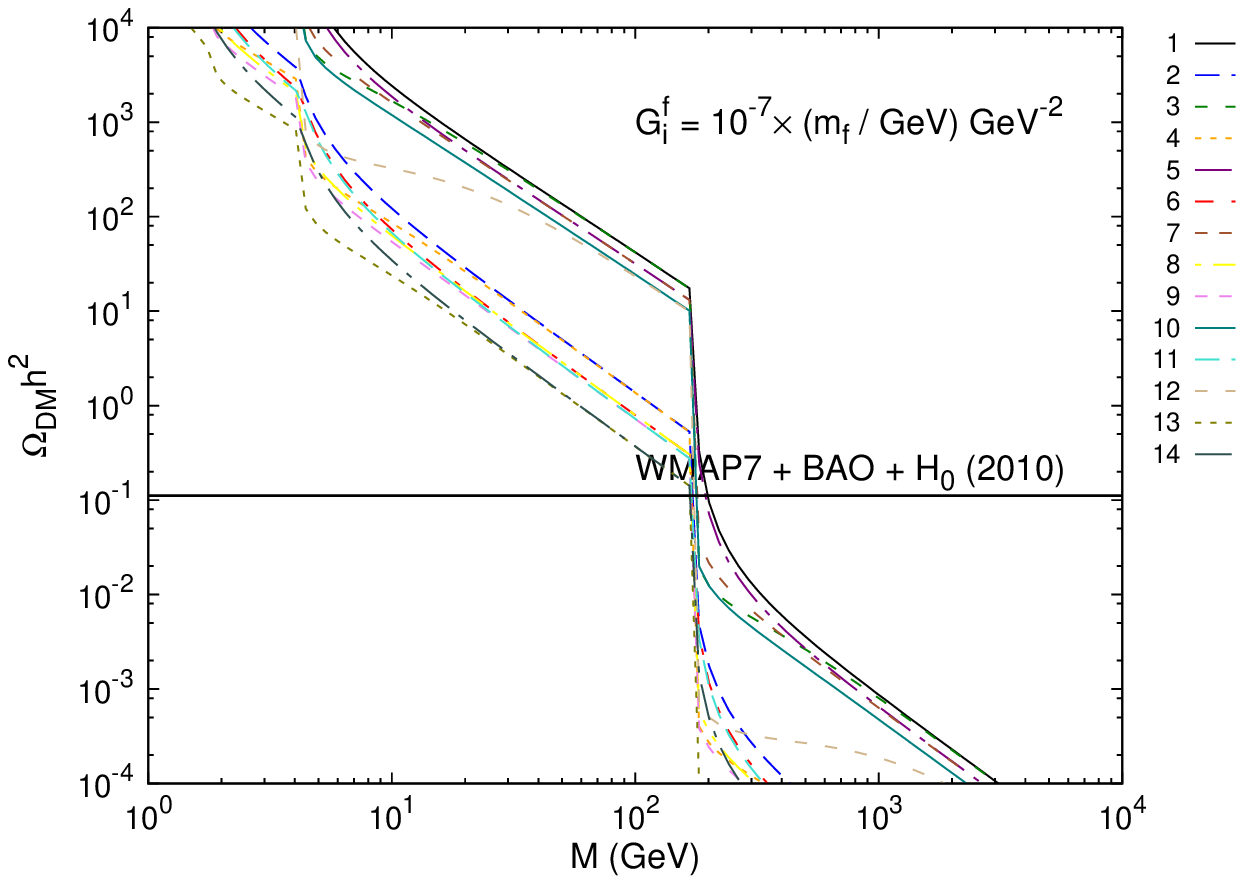}%
\caption{The current relic density (dashed curves) of the spin-3/2
DM, $\Omega_\textrm{DM}h^2$, is predicted as a function of its mass
for various interactions with couplings in scenarios I (left panels)
and II (right panels) respectively. The horizontal solid band shows
the observed range with the best-fit value, $\Omega_{\mathrm{DM}}
h^2=0.1123\pm0.0035$ \cite{Komatsu:2010fb}. The number next to the
legend indicates the operator $\calO_i$.
\label{fig:rd_density}}
\end{figure}

\begin{figure}[!htbp]
\centering
\includegraphics[width=0.49\textwidth]{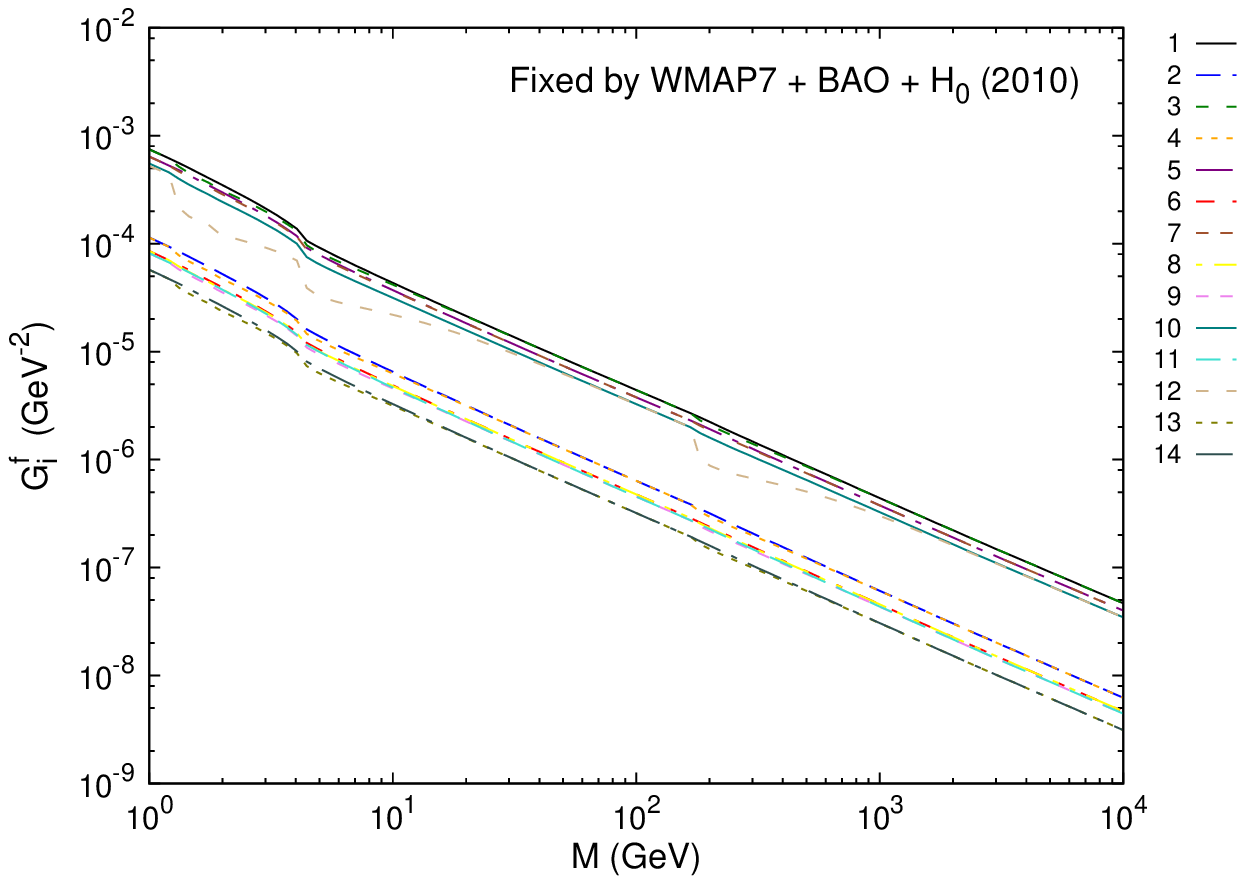}%
\includegraphics[width=0.49\textwidth]{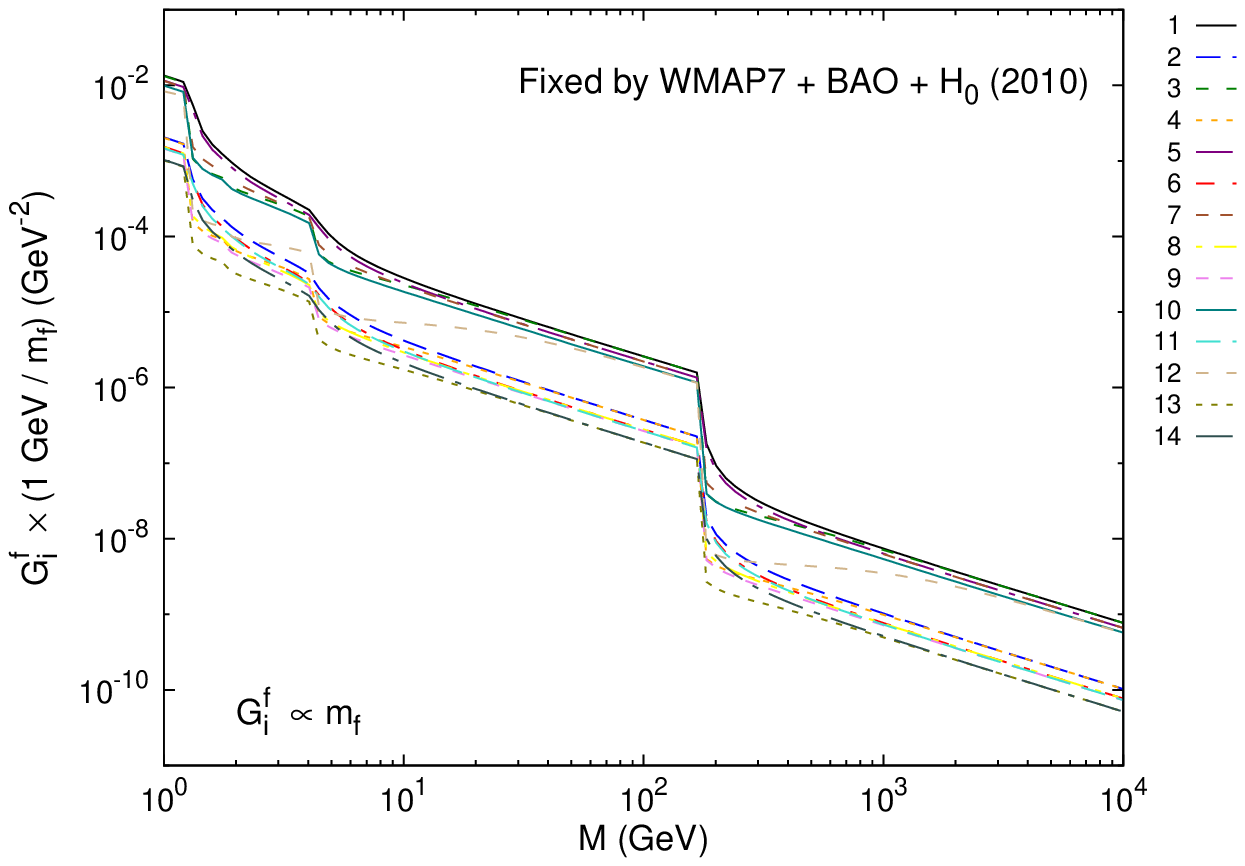}%
\caption{Assuming one of the interactions in eq. (\ref{eq_L})
produces the observed relic density, its coupling is shown as a
function of the DM mass $M$ for both scenarios I (left panel) and II
(right).%
\label{fig:rd_coupling}}
\end{figure}

We show in Fig. \ref{fig:rd_density} the predicted current relic
density for various interactions shown in eq (\ref{eq_L}) as a
function of the mass $M$. For simplicity, we consider two scenarios
for the couplings $G_i^f$ \cite{Beltran:2008xg}. In scenario I, we
assume a universal value for all interactions,
$G_i^f=10^{-5},~10^{-6},~10^{-7}~\GeV^{-2}$. For the operators
involving a chirality-flip bilinear of ordinary fermions, i.e.,
those excluding $\calO_{9,\dots,12}$, it is easy to imagine that
they might be proportional to the mass of the involved fermion. We
therefore study the scenario II in which
$G_i^f(1~\GeV/m_f)=10^{-5},~10^{-6},~10^{-7}~\GeV^{-2}$. For the
purpose of comparison we include the results for
$\calO_{9,\dots,12}$ in scenario II. The predicted density decreases
as the coupling $G_i$ (mass $M$) increases for a fixed mass
(coupling). Also shown (horizontal band) is the range of the
observed DM relic density, corresponding to the best-fit value,
$\Omega_{\mathrm{DM}} h^2=0.1123\pm0.0035$ \cite{Komatsu:2010fb}.
Assuming that one of the interactions in eq (\ref{eq_L}) be
responsible for the observed relic density we plot in Fig.
\ref{fig:rd_coupling} the required couplings as a function of $M$.
Since $\sigma|v|$ increases with $G_i$ and $M$, $G_i$ has to
decrease as $M$ increases in order to match the observed relic
density.

The predicted relic density drops abruptly when a new annihilation
channel is opened with increasing $M$. Similarly, for the relic
density fixed to the observed value the required coupling $G_i$
drops suddenly at each new threshold as $M$ increases. This is
especially obvious in scenario II at the $t\bar t$ threshold where
the effect is significantly enhanced. Most curves fall into one of
the two groups while the one corresponding to the operator
$\calO_{12}^f$ stands alone. This arises from different behavior in
their thermally averaged cross sections,
$\langle\sigma|v|\rangle\approx a+b\langle v^2\rangle$, where the
coefficients $a$ and $b$ correspond to the $s$- and $p$-wave
annihilation respectively. While the operators $\calO_{1,3,5,7,10}$
only give a $b$ term, all others have both $a$ and $b$ terms. In
addition, amongst the latter operators only $\calO_{12}$ has an $a$
term that is proportional to $m_f^2$, which explains its unique
behavior in the figures.

\begin{figure}[!htbp]
\centering
\includegraphics[width=0.44\textwidth]{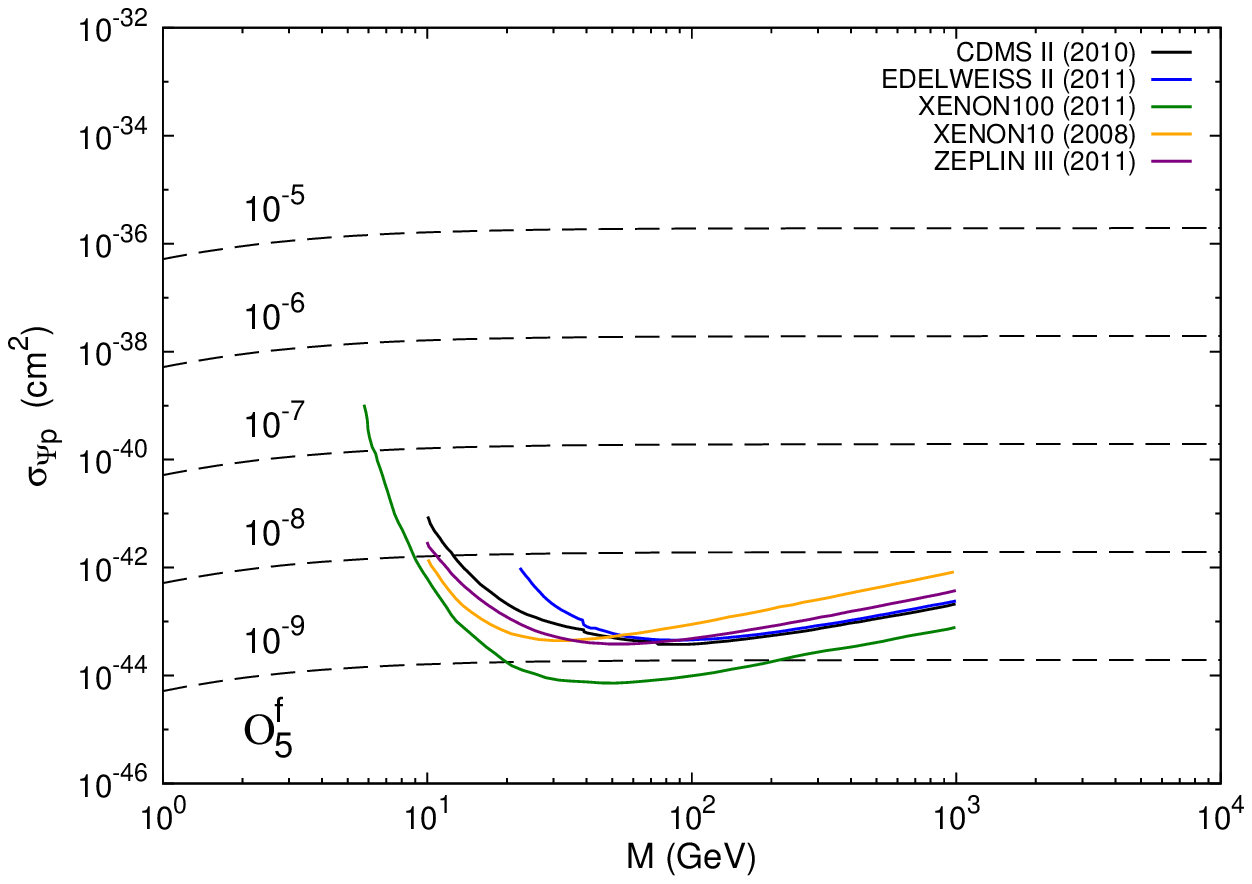}%
\hspace{0.01\textwidth}%
\includegraphics[width=0.44\textwidth]{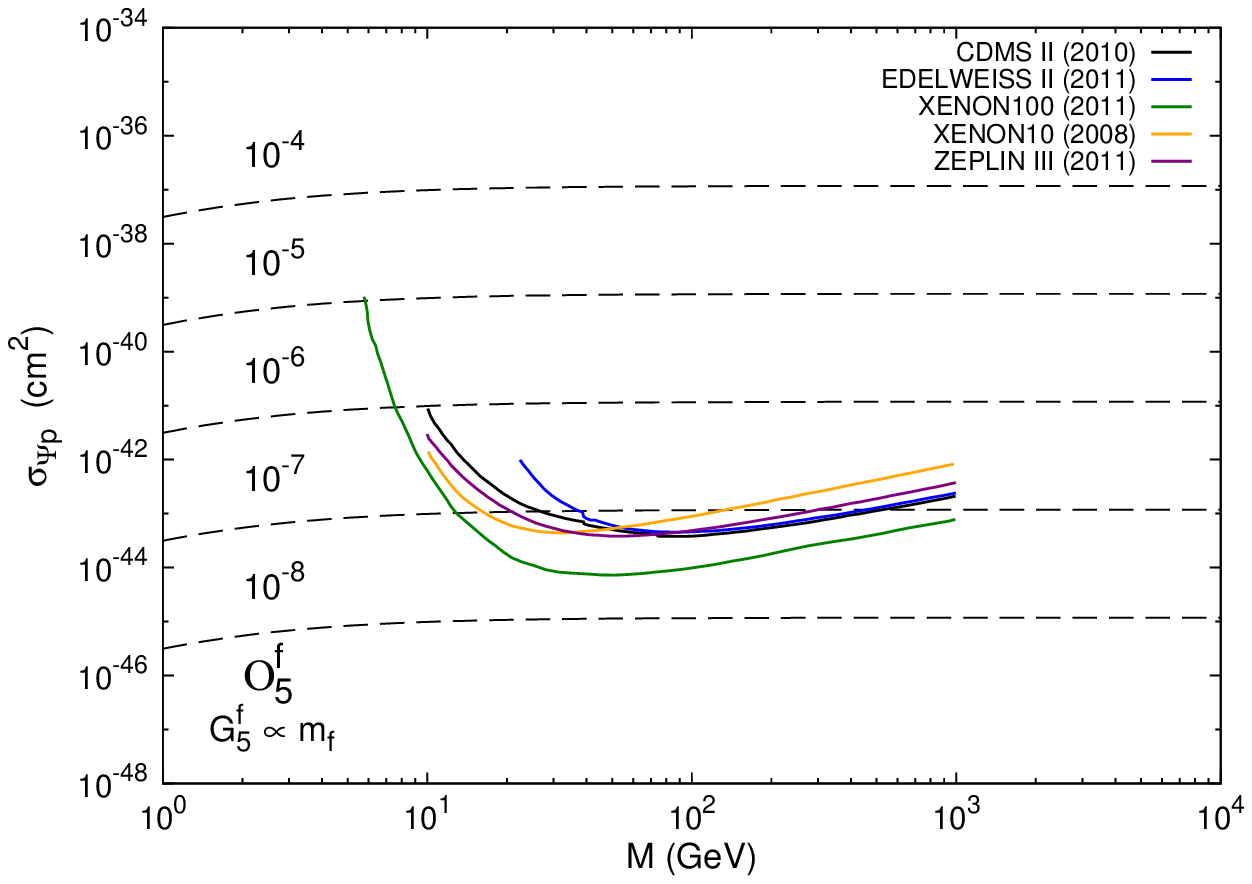}%
\\
\includegraphics[width=0.44\textwidth]{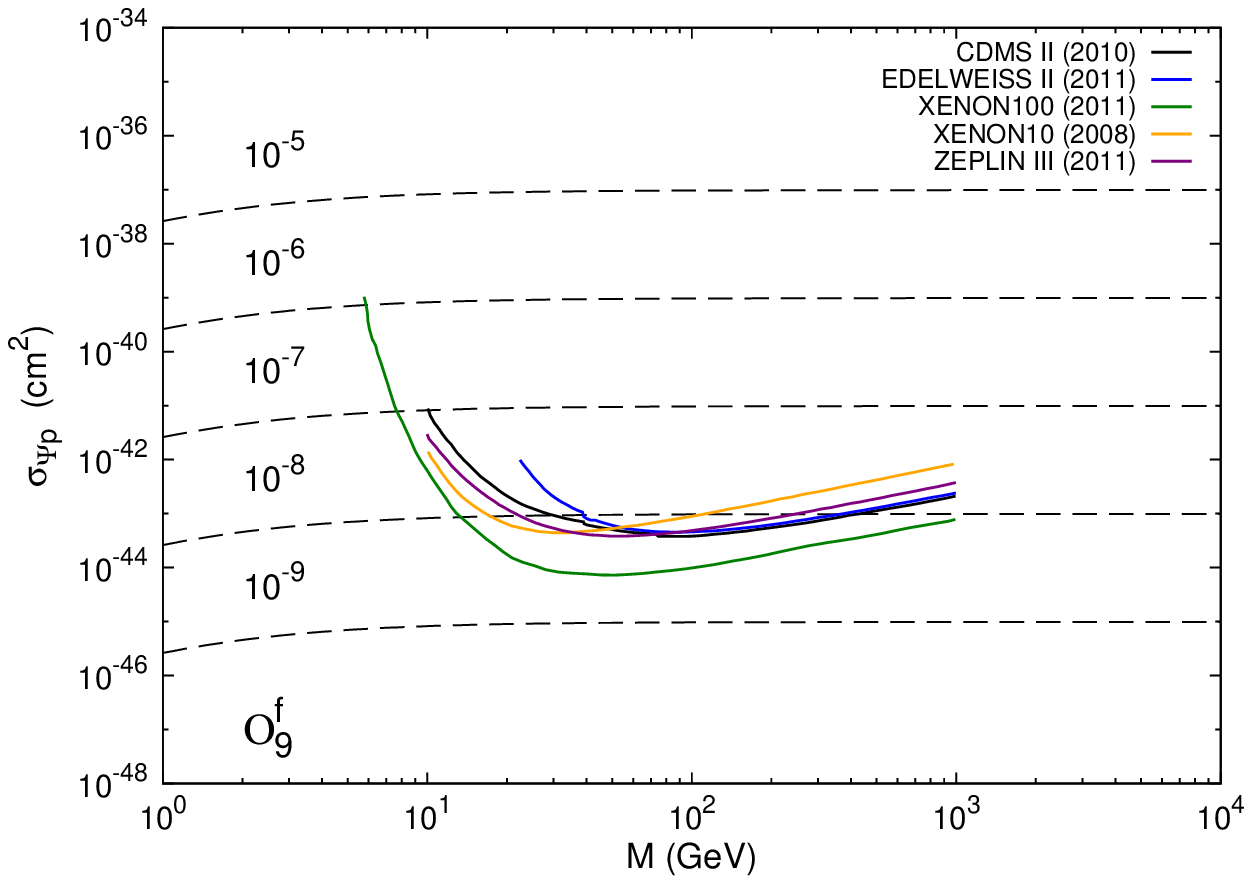}%
\hspace{0.01\textwidth}%
\includegraphics[width=0.44\textwidth]{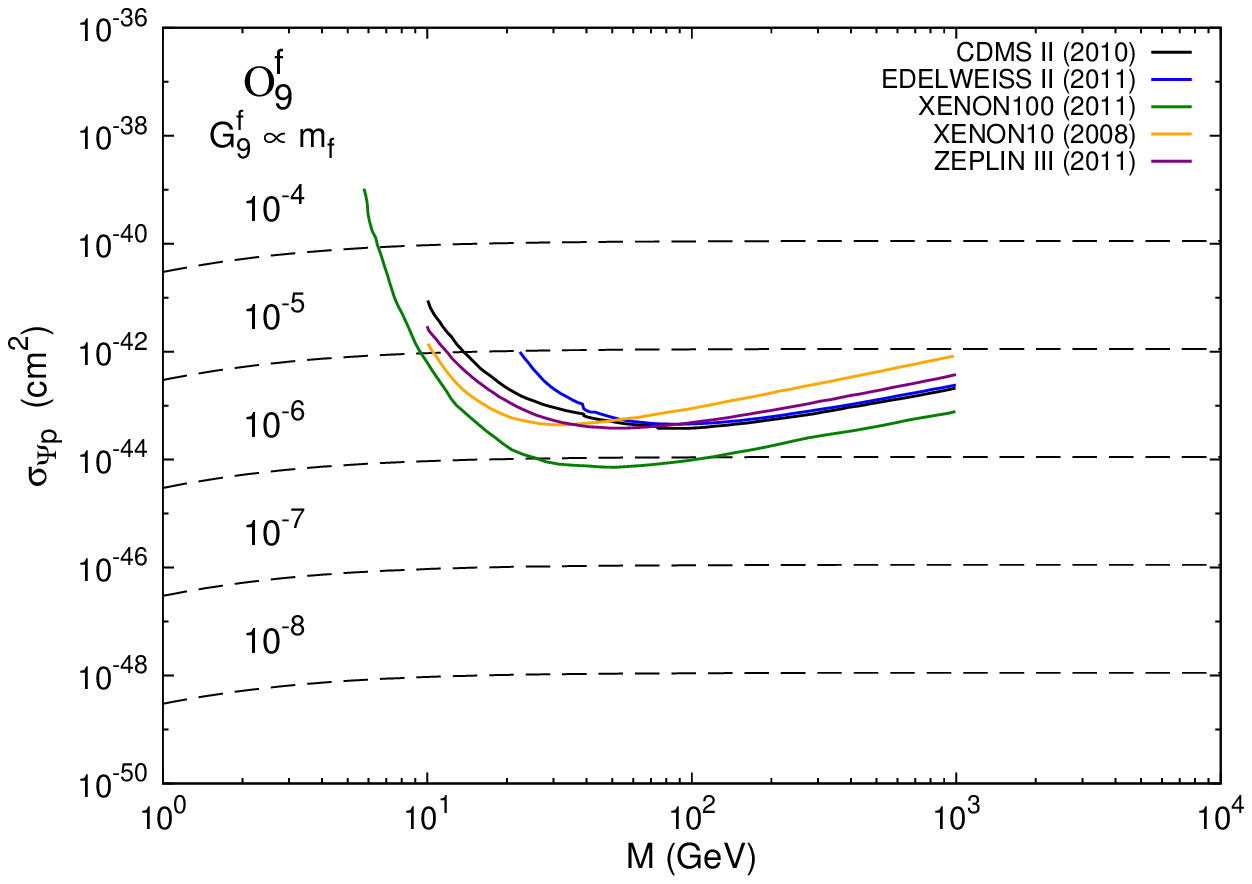}
\caption{The spin-independent $\Psi$-proton cross sections (dashed
curves) are plotted as a function of $M$ at different couplings
for the scalar ($\calO_{5}^{f}$, upper panels) and vector
($\calO_{9}^{f}$, lower panels) interactions. The left panels are
for scenario I with $G_{5,9}^f=10^{-5},\dots,~10^{-8}~\GeV^{-2}$,
while thee right panels are for scenario II with
$G_{5,9}^{f}\times (1~\GeV/
m_f)=10^{-4},~\dots,~10^{-8}~\GeV^{-2}$. The solid curves are the
upper bounds from the experiments CDMS II (2010)
\cite{Ahmed:2009zw}, EDELWEISS-II (2011) \cite{Armengaud:2011},
XENON100 (2011) \cite{Aprile:2011hi}, XENON10 (2008)
\cite{Angle:2007uj}, and ZEPLIN-III
(2011) \cite{Akimov:2011tj}.%
\label{fig:scat:SI}}
\end{figure}

\subsection{Direct detection\label{sec-direct}}

The experimental results in direct detection of dark matter are
conventionally presented in terms of cross sections on nucleons. We
follow this practice to show our results on the cross sections
$\sigma_{\Psi p(n)}^i$ for nonrelativistic scattering of $\Psi$ off
a proton (neutron) due to various operators $\calO_i^q$. As we
explained in the last section, out of many possible interactions
there are only two types of them that contribute to the SI cross
section and three types to the SD one. $\sigma_{\Psi p(n)}^i$ are
still given by eqs (\ref{eq_SIs},\ref{eq_SIv},\ref{eq_SD}) for the
SI and SD cases respectively, with the following substitutions:
$f_N^5\to f_{p(n)}^5$, $b_N^9\to b_{p(n)}^9$,
$g_N^z\to\sum_qG_z^q\Delta_q^{p(n)}$, $J_N\to 1/2$, and $\mu\to
Mm_{p(n)}/(M+m_{p(n)})$. For the chiral parameters related to the
scalar SI matrix element we use the values in Ref.
\cite{Ellis:2000ds}: $f_{T_{u}}^{(p)}=0.020\pm0.004$,
$f_{T_{d}}^{(p)}=0.026\pm0.005$, $f_{T_{s}}^{(p)}=0.118\pm0.062$ for
the proton, and $f_{T_{u}}^{(n)}=0.014\pm0.003$, $f_{T_{d}}^{(n)}
=0.036\pm0.008$, $f_{T_{s}}^{(n)}=0.118\pm0.062$ for the neutron.
For the nucleon spin fractions carried by quarks that are required
in the SD matrix element, we assume the values in Ref.
\cite{Belanger:2008sj}: $\Delta_{u}^{p}=\Delta_{d}^{n}=0.78\pm0.02$,
$\Delta_{d}^{p} =\Delta_{u}^{n}=-0.48\pm0.02$, and
$\Delta_{s}^{p}=\Delta_{s}^{n}=-0.15\pm0.02.$

\begin{figure}[!htbp]
\centering
\includegraphics[width=0.44\textwidth]{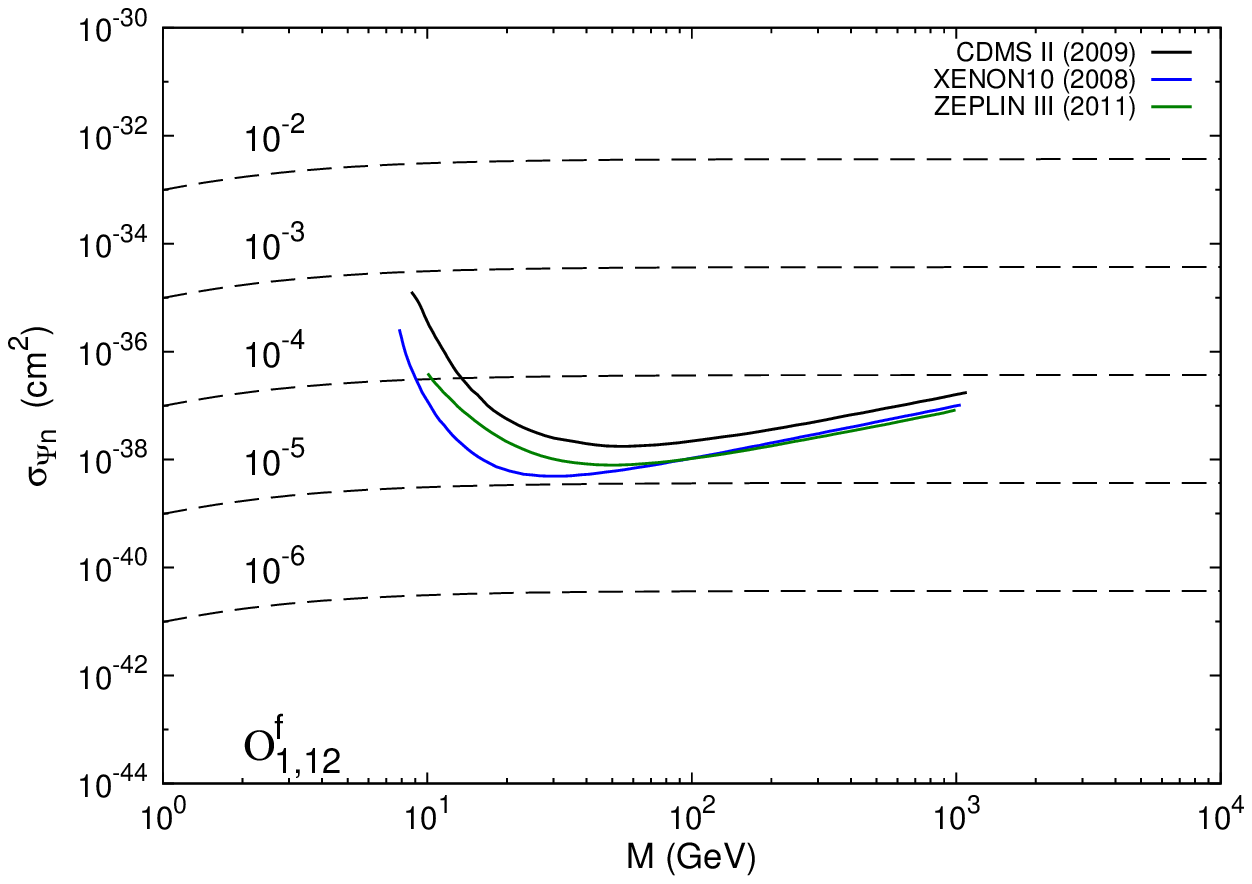}%
\hspace{0.01\textwidth}%
\includegraphics[width=0.44\textwidth]{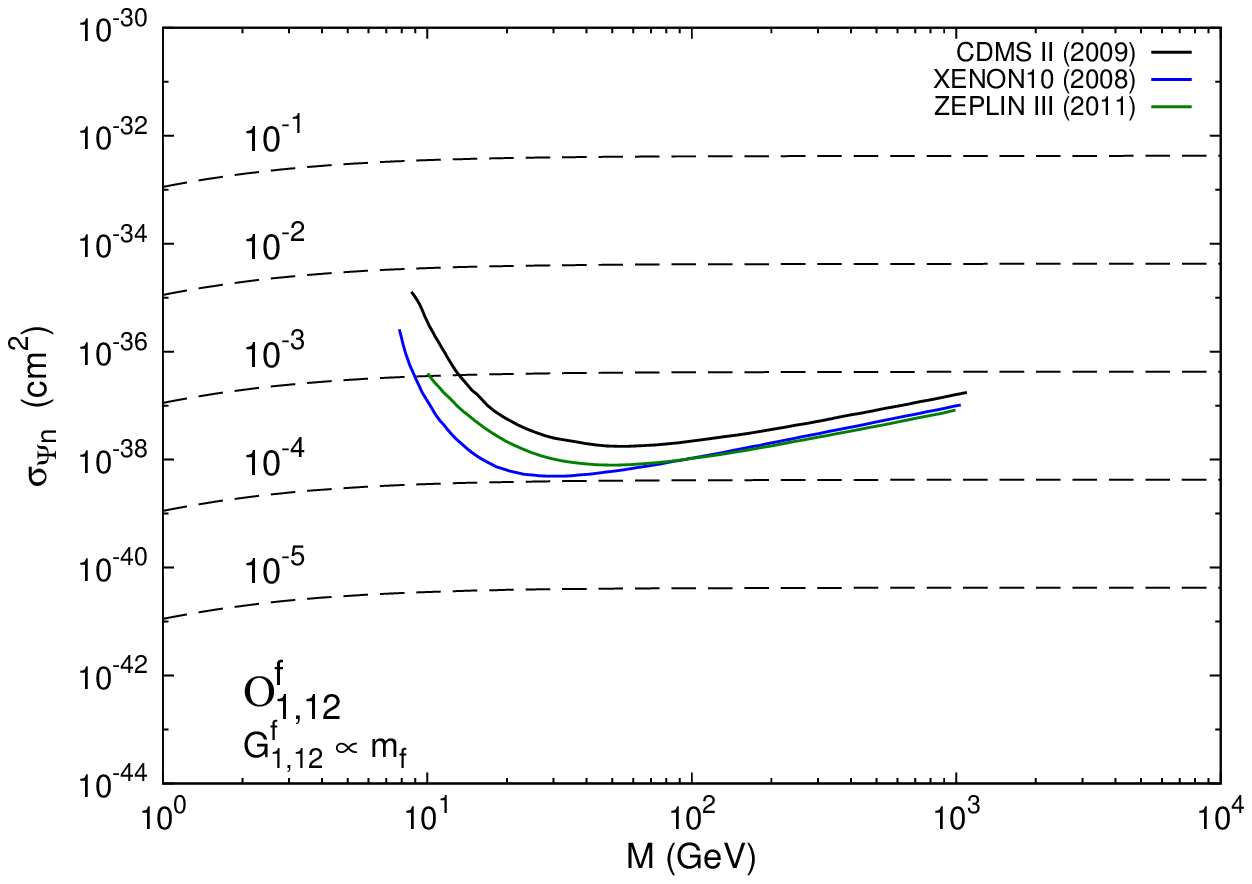}
\\
\includegraphics[width=0.44\textwidth]{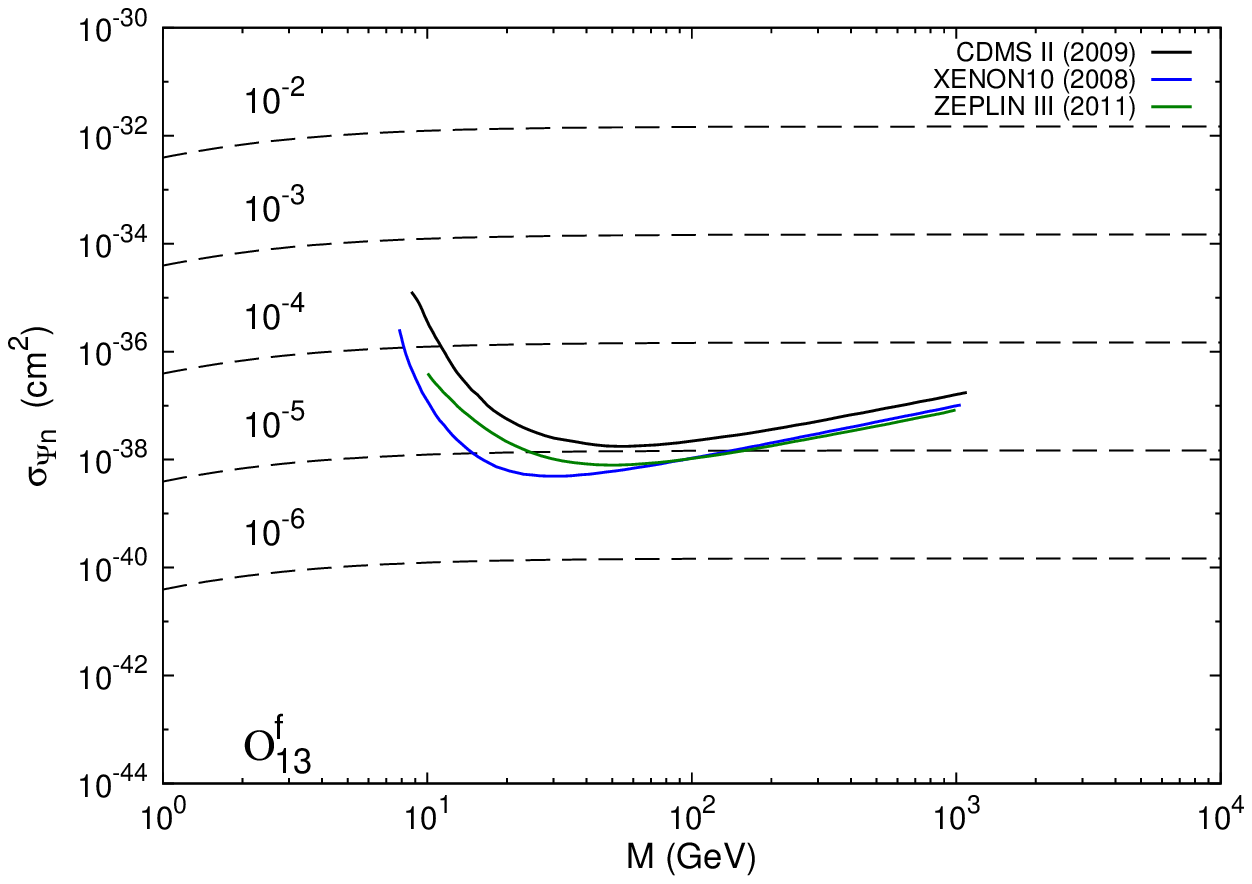}%
\hspace{0.01\textwidth}%
\includegraphics[width=0.44\textwidth]{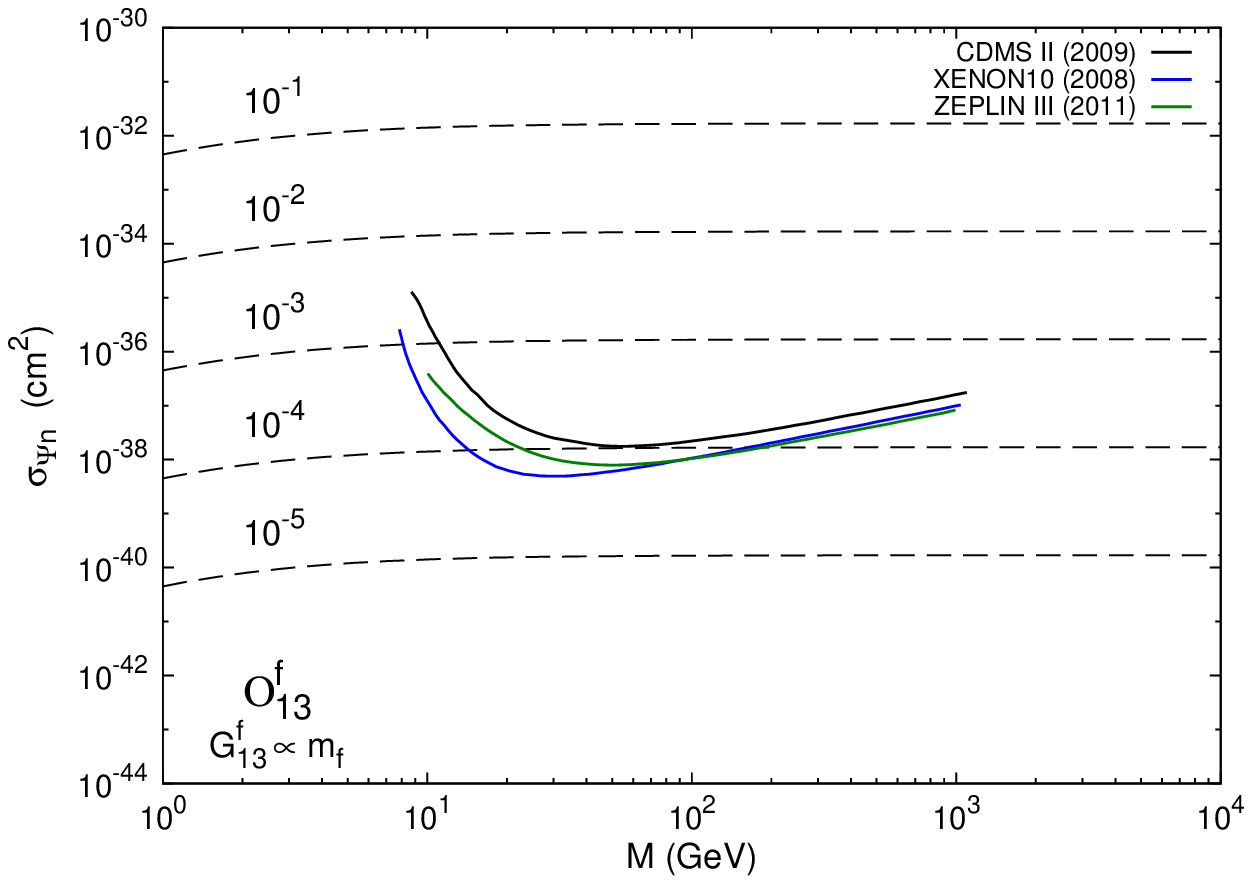}
\caption{The spin-dependent $\Psi$-neutron cross sections (dashed
curves) are plotted as a function of $M$ for the axial vector
$\calO_{12}^f$ and tensor $\calO_{1,13}^f$ interactions. The solid
curves are the upper bounds from the experiments CDMS II(2009)
\cite{Ahmed:2009}, XENON10 (2008) \cite{Angle:2008we} and ZEPLIN-III
(2011) \cite{Akimov:2011tj}. The notations for couplings are similar
to Fig. \ref{fig:scat:SI}.
%
%
\label{fig:dirac:scat:SD}}
\end{figure}

In Fig. \ref{fig:scat:SI} we display the SI $\Psi$-proton cross
sections $\sigma_{\Psi p}^{5,9}$ as a function of the mass $M$ in
both scenarios of couplings, while in Fig. \ref{fig:dirac:scat:SD}
we plot the SD $\Psi$-neutron cross sections $\sigma_{\Psi
n}^1=\sigma_{\Psi n}^{12}$ and $\sigma_{\Psi n}^{13}$. In scenario
II, both light and heavy quarks contribute equally to the scalar SI
$\sigma^5_{\Psi p(n)}$ while for the vector SI $\sigma^5_{\Psi
p(n)}$ both contributions are suppressed by either a light quark
mass or a vanishingly small content of heavy quarks in the nucleon.
In the same scenario, the contributions to the SD
$\sigma^{1,12,13}_{\Psi p(n)}$ from both light and heavy quarks are
significantly suppressed by either light quark masses or tiny spin
fractions of heavy quarks in the nucleon. In scenario I with
universal couplings both SI and SD cross sections are dominated by
light quarks. Since the current experimental bounds on the SD cross
sections are several orders of magnitude weaker than the SI ones,
the upper bounds that one can get on the SD effective couplings are
also much weaker.

\subsection{Indirect detection\label{sec-indirect}}

The DM particles in our Galaxy can annihilate through the
interactions in eq (\ref{eq_L}) to produce leptons and quarks that
fragment and interact further with the interstellar gas to create
more secondaries, including gamma rays, neutrinos, positrons and
antiprotons. By comparing the observed cosmic ray fluxes with known
astrophysical sources, it is possible to infer the properties of DM
particles in our galactic halo and constrain their annihilation
rate. We have employed the public computer code GALPROP
\cite{Strong:1998pw} to simulate the antiproton to proton flux
ratio. The code solves numerically with appropriate boundary
conditions the transport equation for the number density of cosmic
particles that takes into account diffusion and convection effects
amongst others. For our purpose here, the source term in the
equation will contain a piece due to the $\Psi\bar\Psi$ annihilation
\begin{eqnarray}
Q_\Psi^{\bar p}(r,E)&=&\bigg[\frac{\rho(r)}{2M}\bigg]^2\sum_q
\langle\sigma^q|v|\rangle\bigg(\frac{dN}{dE}\bigg)_q^{\bar p},
\end{eqnarray}
where the sum is over all channels of quark production,
$(dN/dE)_q^{\bar p}$ is the antiproton number per unit energy
produced in the $q\bar q$ channel, and $\rho(r)$ is the mass
density distribution of the DM particles. We use the Monte-Carlo
program PYTHIA \cite{Sjostrand:2006za} to simulate the
$(dN/dE)_q^{\bar p}$ spectrum.

In our numerical analysis, we take the NFW profile
\cite{Navarro:1995iw}:
\begin{eqnarray}
\frac{\rho(r)}{\rho_{\odot}}=\frac{r_{\odot}}{r}
\left[\frac{1+r_{\odot}/R}{1+r/R}\right]^{2},
\end{eqnarray}
where $\rho_{\odot}$ is the DM density at the solar location,
$r_{\odot}$ the distance of the sun to the Galactic center, and $R$
the scale radius. We adopt the following values for these parameters
from Table 3 in Ref.\cite{Riccardo:2010} :
$\rho_{\odot}=0.389~\GeV~\cm^{-3}$, $r_{\odot}=8.28~\kpc$, and
$R=20~\kpc$. The Galactic DM particles should follow the
Maxwell-Boltzmann velocity distribution. We choose the velocity
dispersion $\bar v\equiv\sqrt{\langle
v^2(r_{\odot})\rangle}=\sqrt{3/2}v_c(r_{\odot})$ with
$v_c(r_{\odot})=243.75\rm{km}\rm{s}^{-1}$ being the local circular
velocity \cite{Riccardo:2010}, so that $\langle v^2\rangle=2\langle
v^2(r_{\odot})\rangle$.

In the calculation of the $\bar p/p$ flux ratio with GALPROP, the
diffusion region of cosmic rays is described by a thick disk of
thickness $2L\approx 8~\kpc$ and radius $R\approx 20~\kpc$, with the
thin galactic disk of thickness $2h\approx 200~\pc$ and radius R
lying in the middle. The charged particles traversing the solar
system are affected by the solar wind, which results in a shift in
the spectrum observed at the Earth compared to the interstellar one
\cite{Gleeson:1968zz,Belanger:2010gh}. We have scanned the solar
modulation potential $\Phi$ from $300$ to $1000~\textrm{MV}$, and
found that $\Phi=330~\textrm{MV}$ yields the minimal $\chi^2$ for
the background flux.

\begin{figure}[!htbp]
\centering
\includegraphics[width=0.49\textwidth]{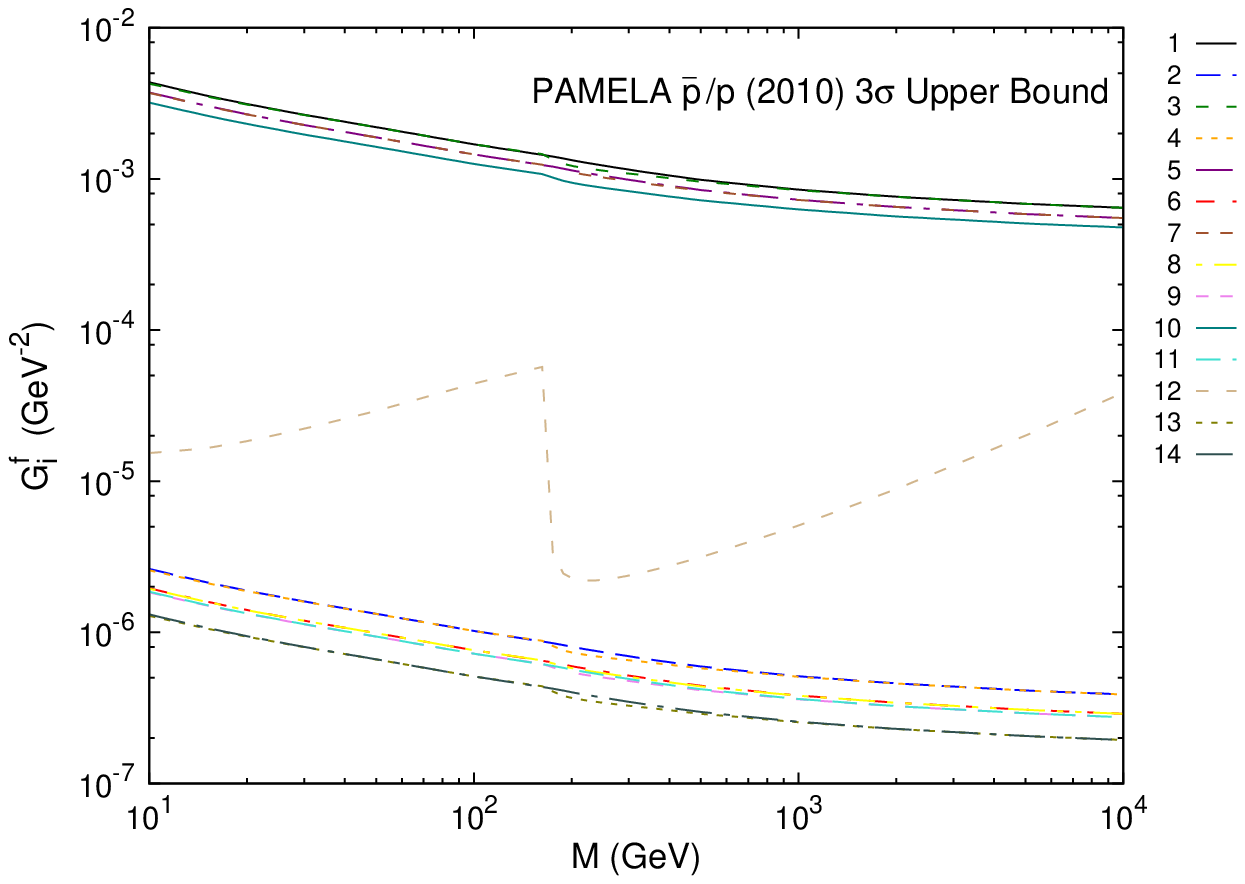}%
\includegraphics[width=0.49\textwidth]{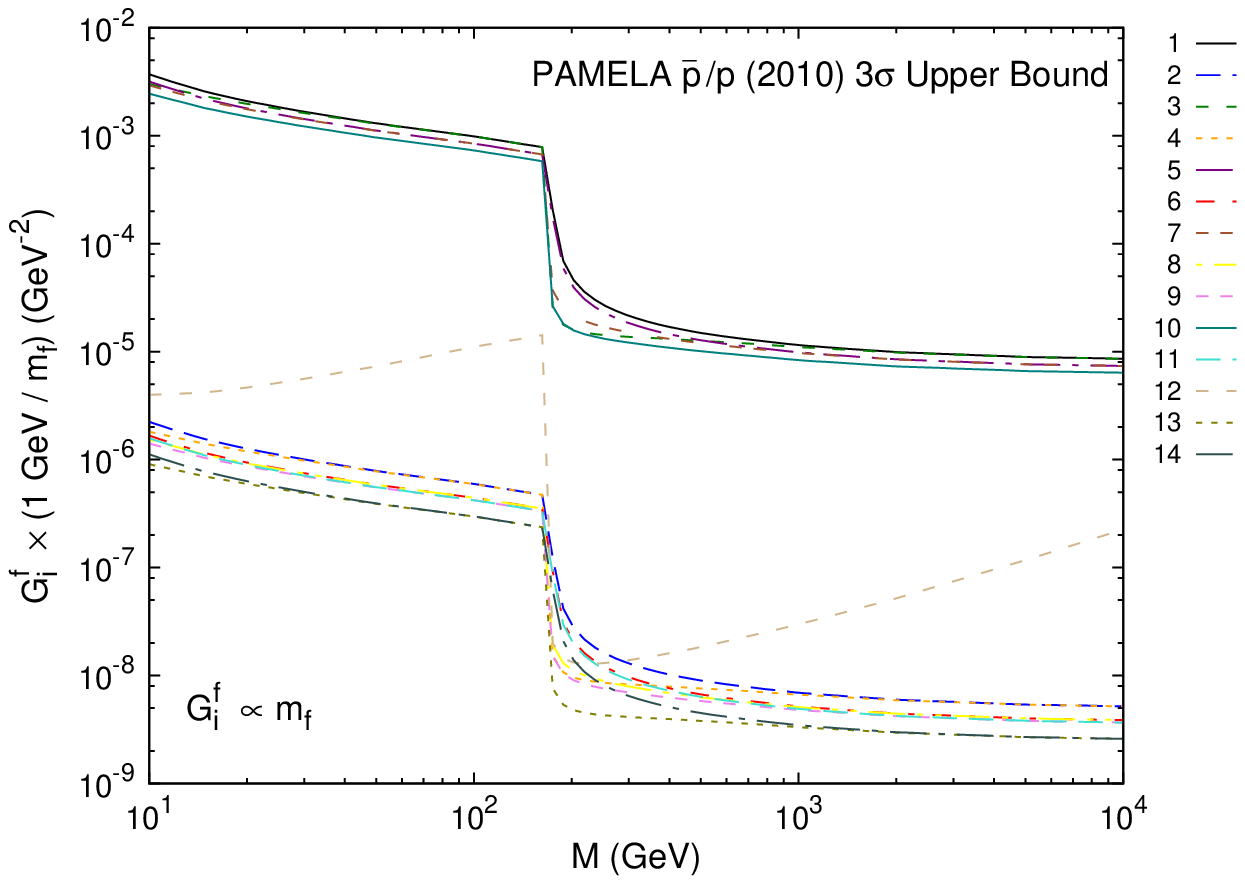}%
\caption{The $3\sigma$ upper bounds on various couplings $G_i^f$ as
a function of $M$ as imposed by the PAMELA $\bar p / p$ spectrum
\cite{Adriani:2010rc}, in scenarios I (left panel) and II (right).
\label{fig-antiproton}}
\end{figure}

Although the PAMELA data on the $\bar p/p$ flux ratio can be
accounted for by GALPROP based on the conventional propagation model
of cosmic rays, it cannot exclude a small portion of contribution
from DM annihilations. This will set a stringent bound on the
annihilation cross section $\langle\sigma |v|\rangle$. By varying it
within the acceptable deviation ranges of the PAMELA data and
evaluating the $\chi^2$ value, we obtain the $3\sigma$ upper bounds
on the couplings $G_{i}^{f}$ for a given value of the mass $M$. The
results are shown in Fig. \ref{fig-antiproton} for both scenarios I
and II. The behavior of the curves is quite similar to that shown in
Fig. \ref{fig:rd_coupling}, but the drop is less steep as $M$
increases.

We do not consider here the PAMELA positron fraction excess and
related effective interactions for a few reasons. It has been
proposed that the excess could originate from some astrophysical
sources such as supernova remnants or nearby pulsars that were not
accounted for earlier, see for instance Ref. \cite{Cirelli:2012tf}
for a status review. If the excess is due mainly to the dark
matter annihilation, a strong tension arises between the excess
and the relic density that is tentatively parameterized by a
`boost factor' as large as a few hundreds or even a thousand,
whose origin however is unclear \cite{He:2009ra}. Thus the excess
itself cannot yet result in useful constraints on interactions
with leptons. And finally we want to work out combined constraints
in the next section where only the interactions with quarks are
relevant in direct detections.

\subsection{Combined constraints\label{sec-combine}}

\begin{figure}[!htbp]
\centering
\includegraphics[width=0.33\textwidth]{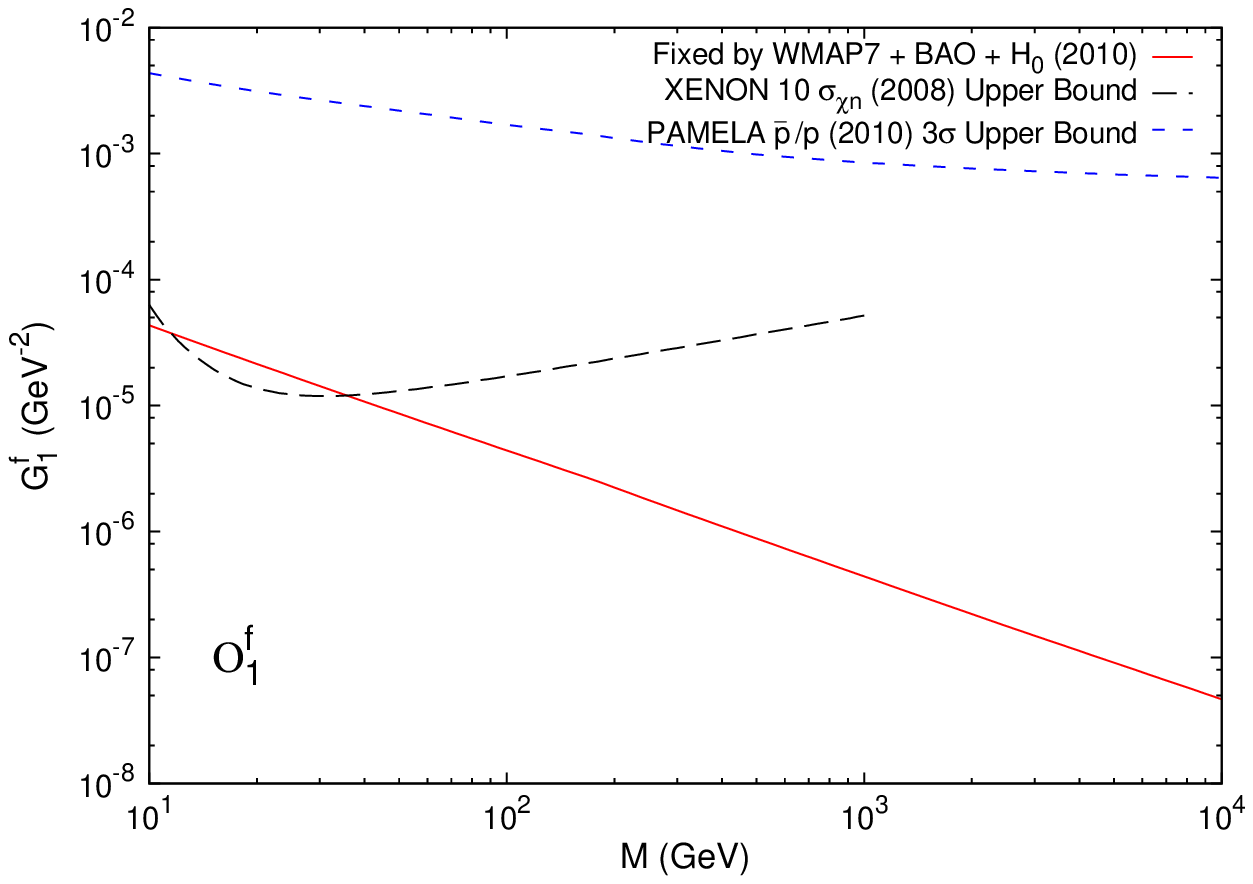}%
\hspace{0.008\textwidth}%
\includegraphics[width=0.33\textwidth]{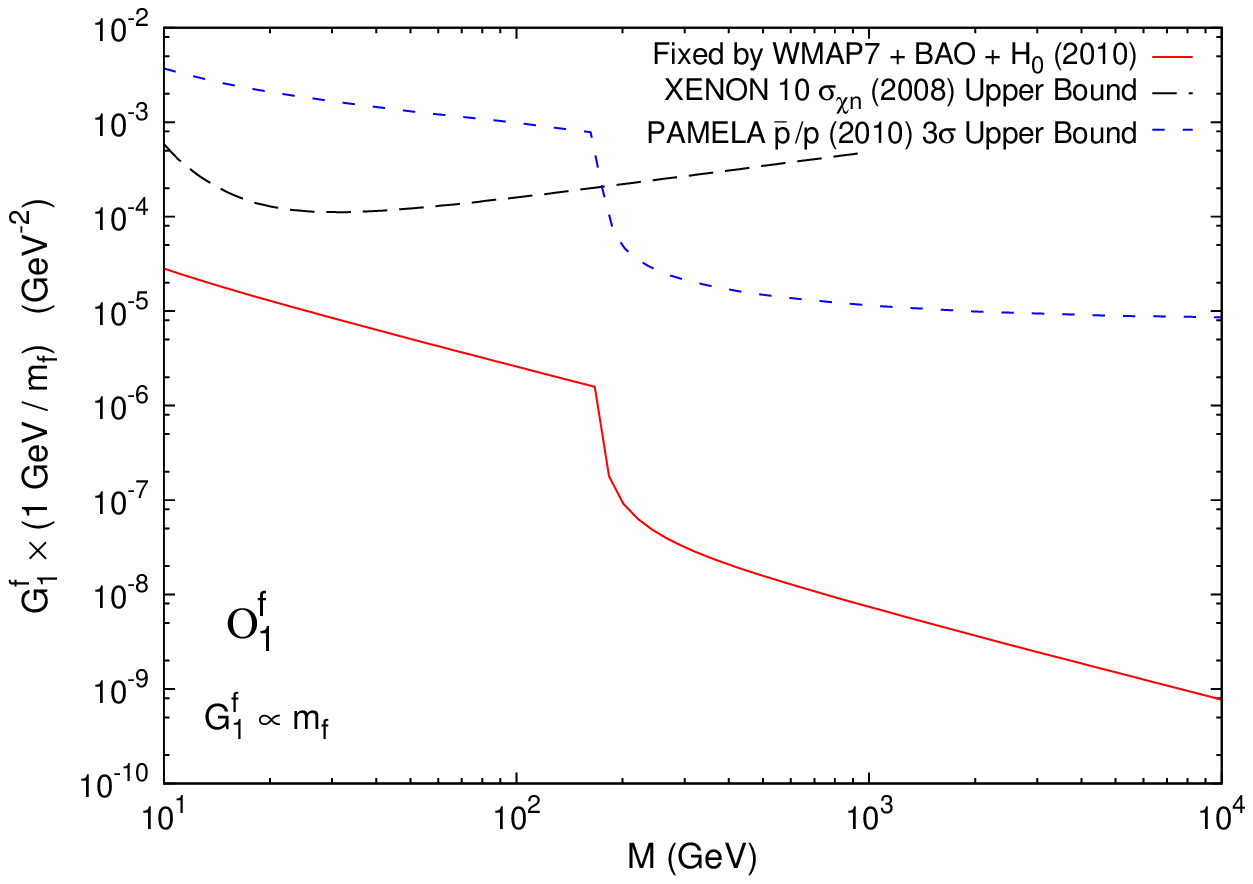}%
\\
\includegraphics[width=0.33\textwidth]{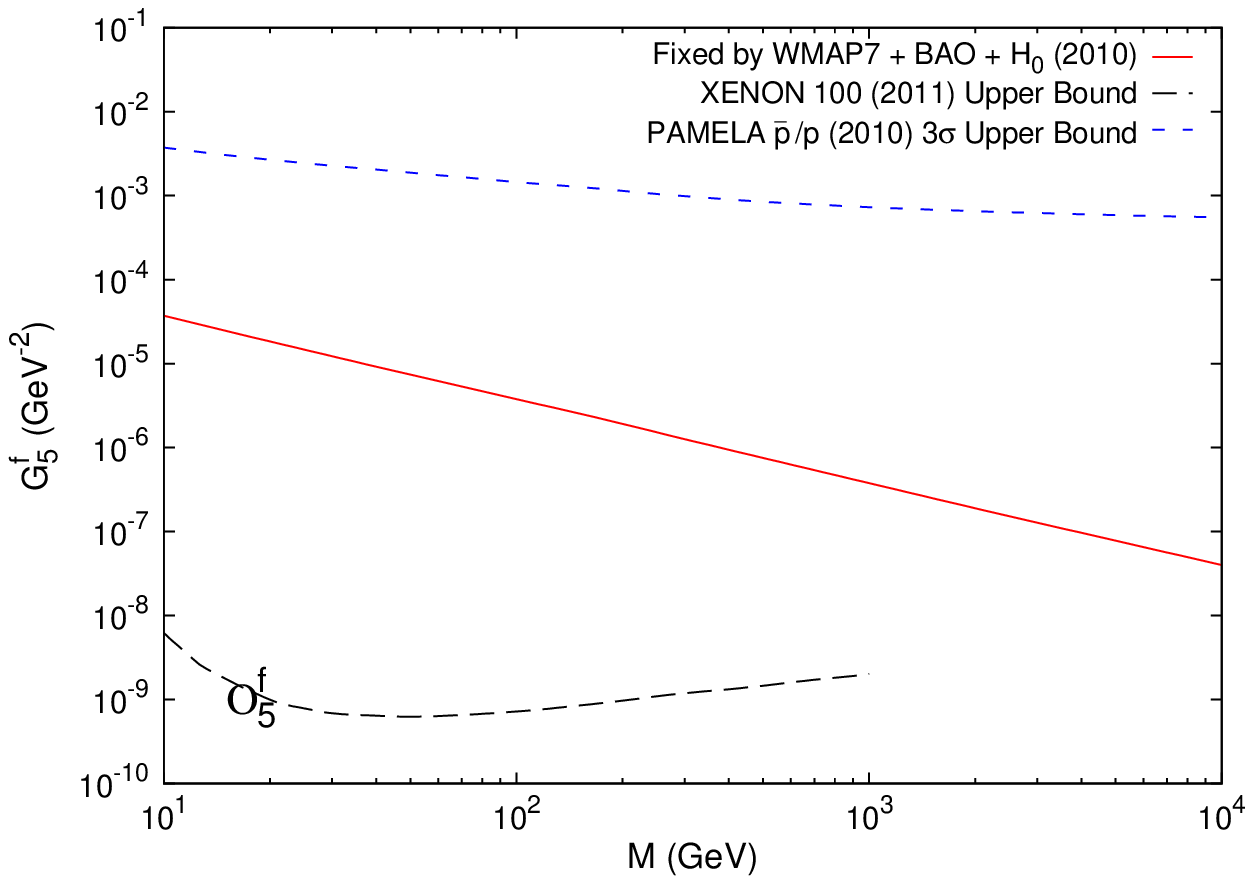}%
\hspace{0.008\textwidth}%
\includegraphics[width=0.33\textwidth]{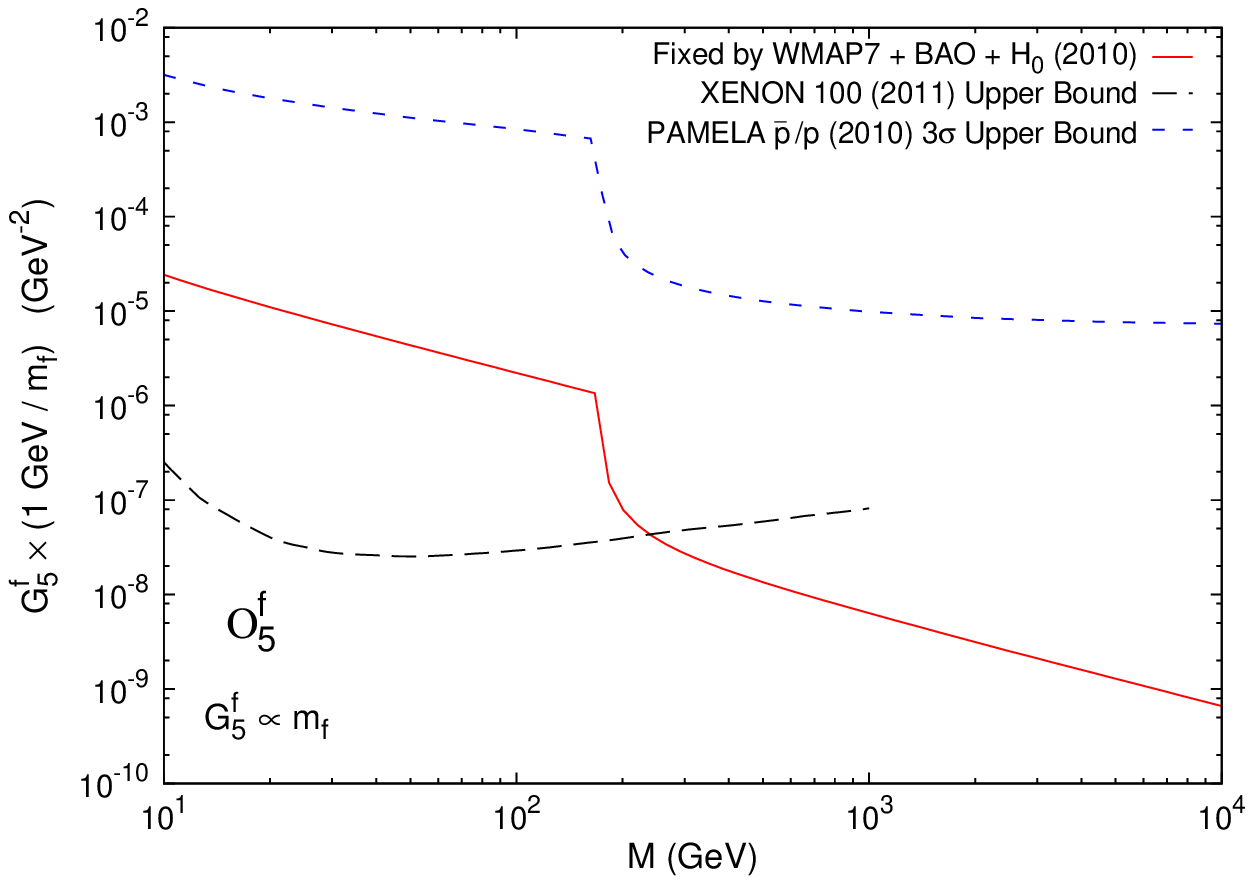}%
\\
\includegraphics[width=0.33\textwidth]{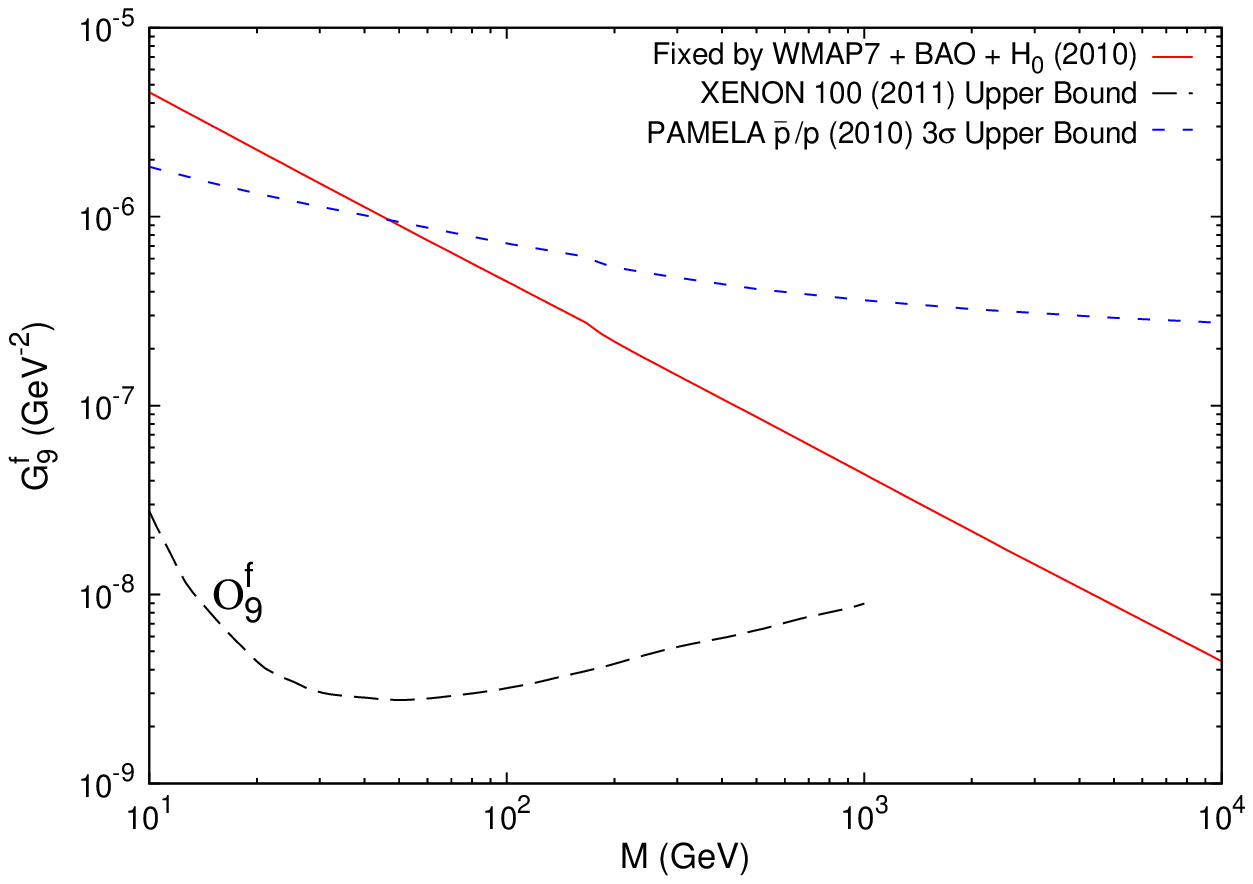}%
\hspace{0.008\textwidth}%
\includegraphics[width=0.33\textwidth]{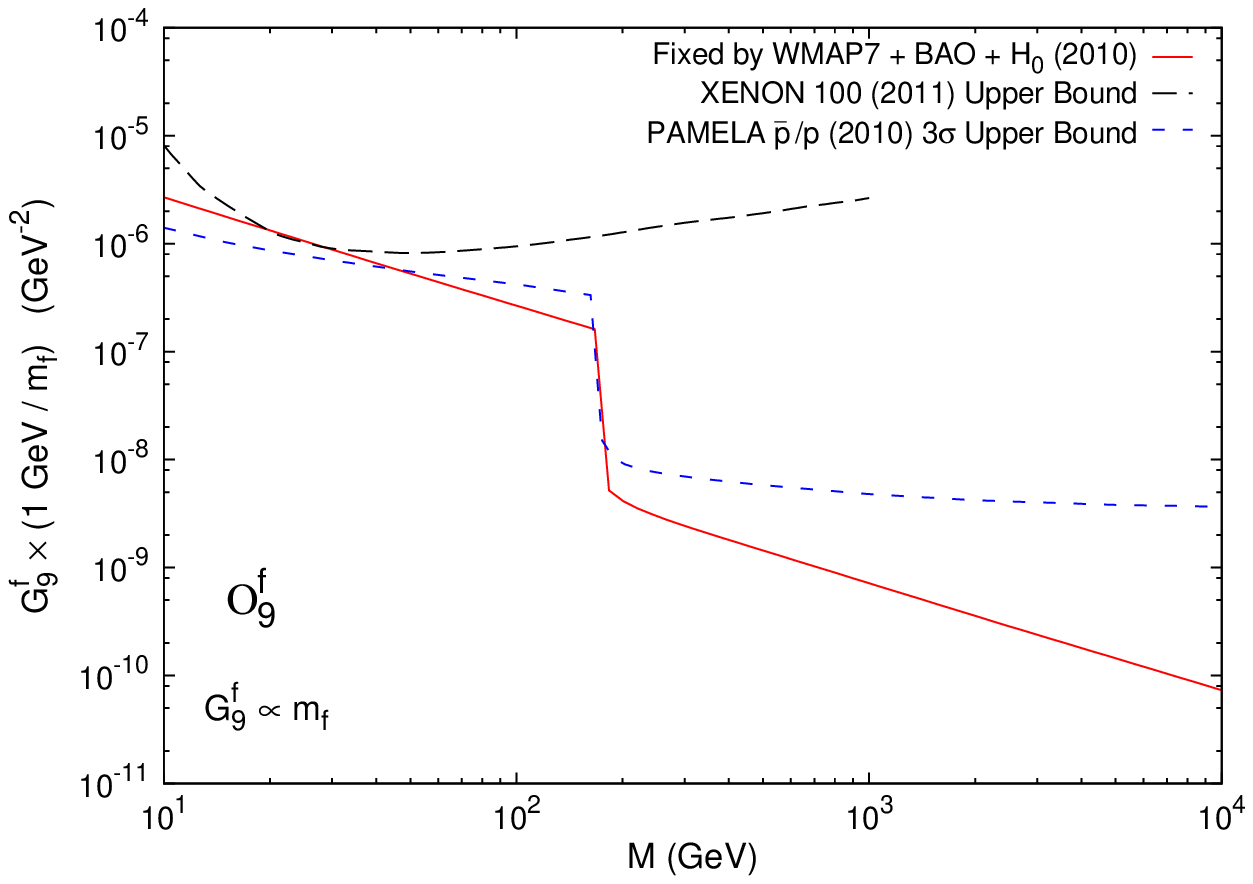}%
\\
\includegraphics[width=0.33\textwidth]{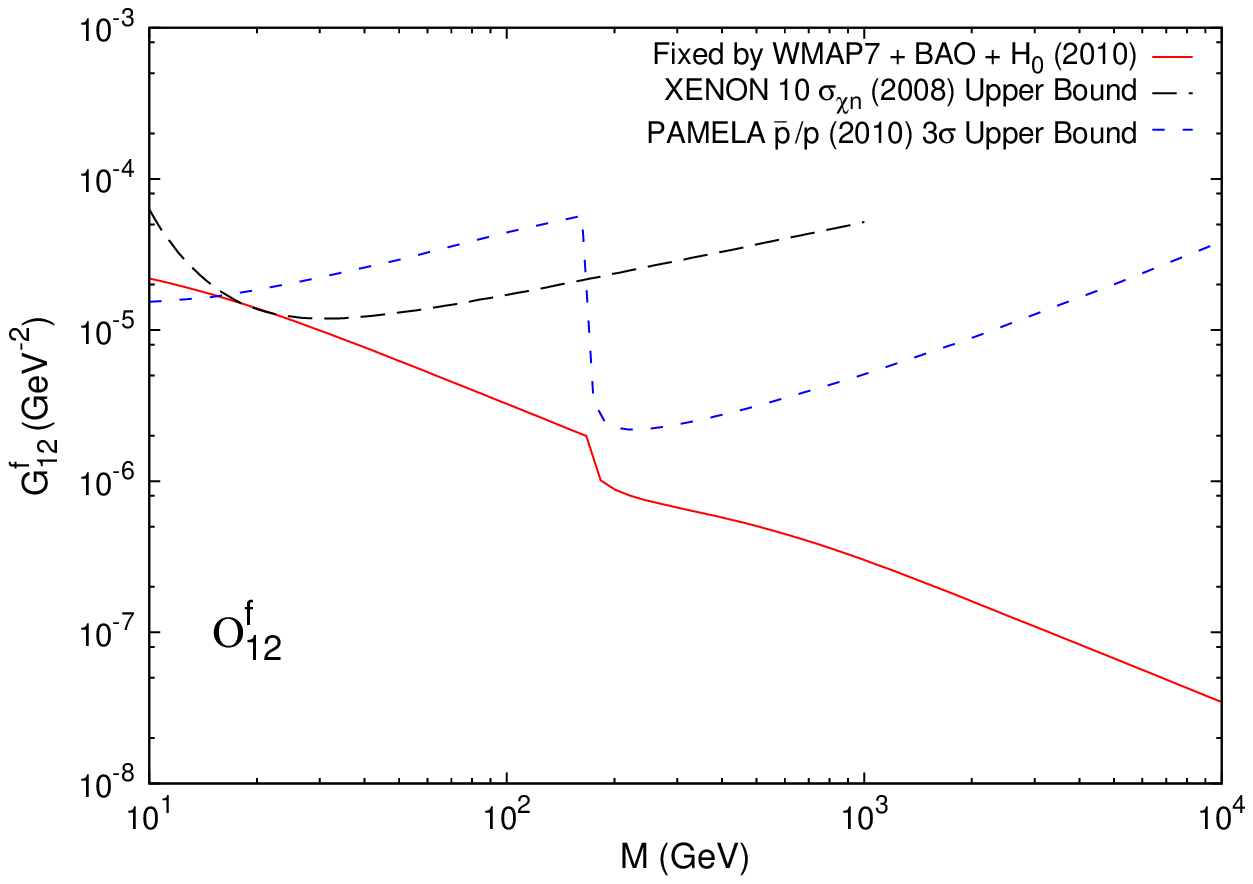}%
\hspace{0.008\textwidth}%
\includegraphics[width=0.33\textwidth]{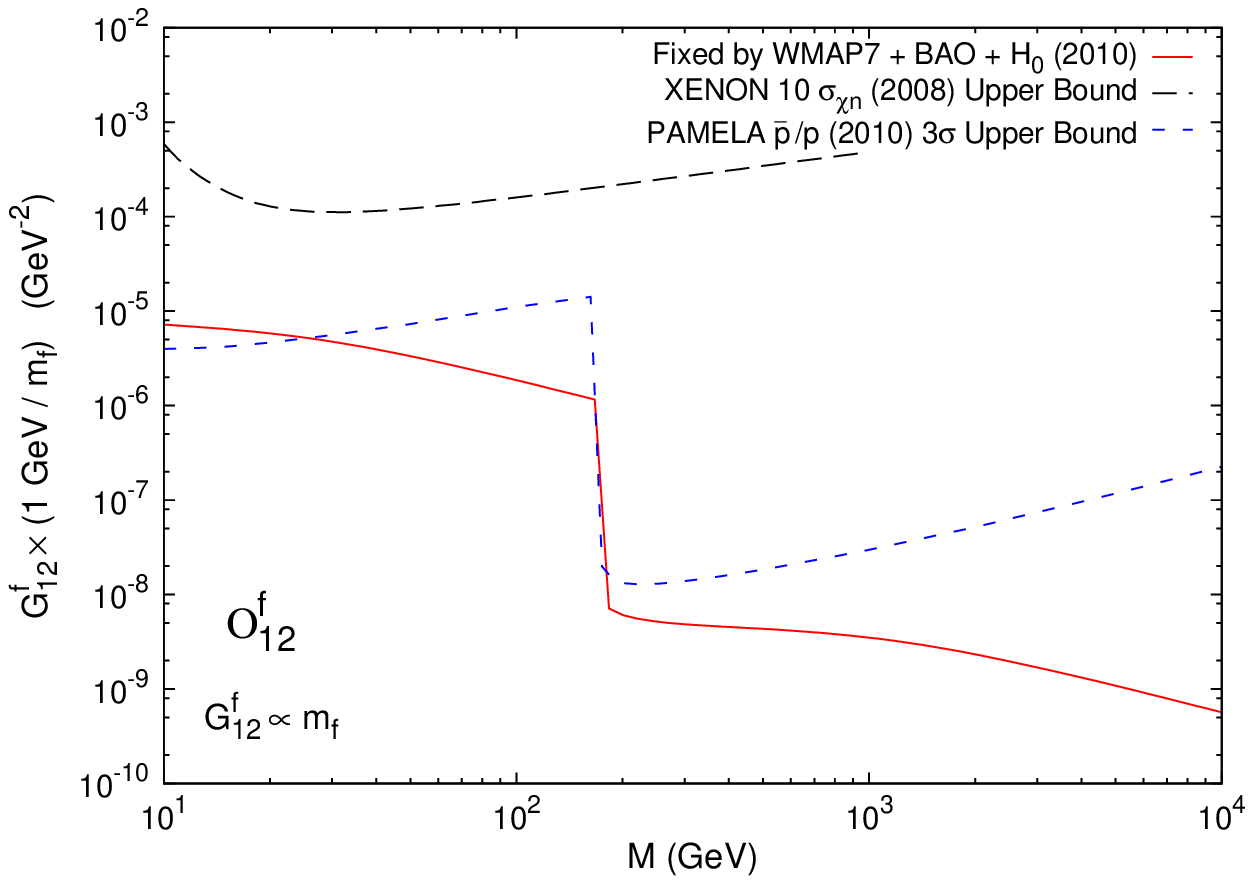}%
\\
\includegraphics[width=0.33\textwidth]{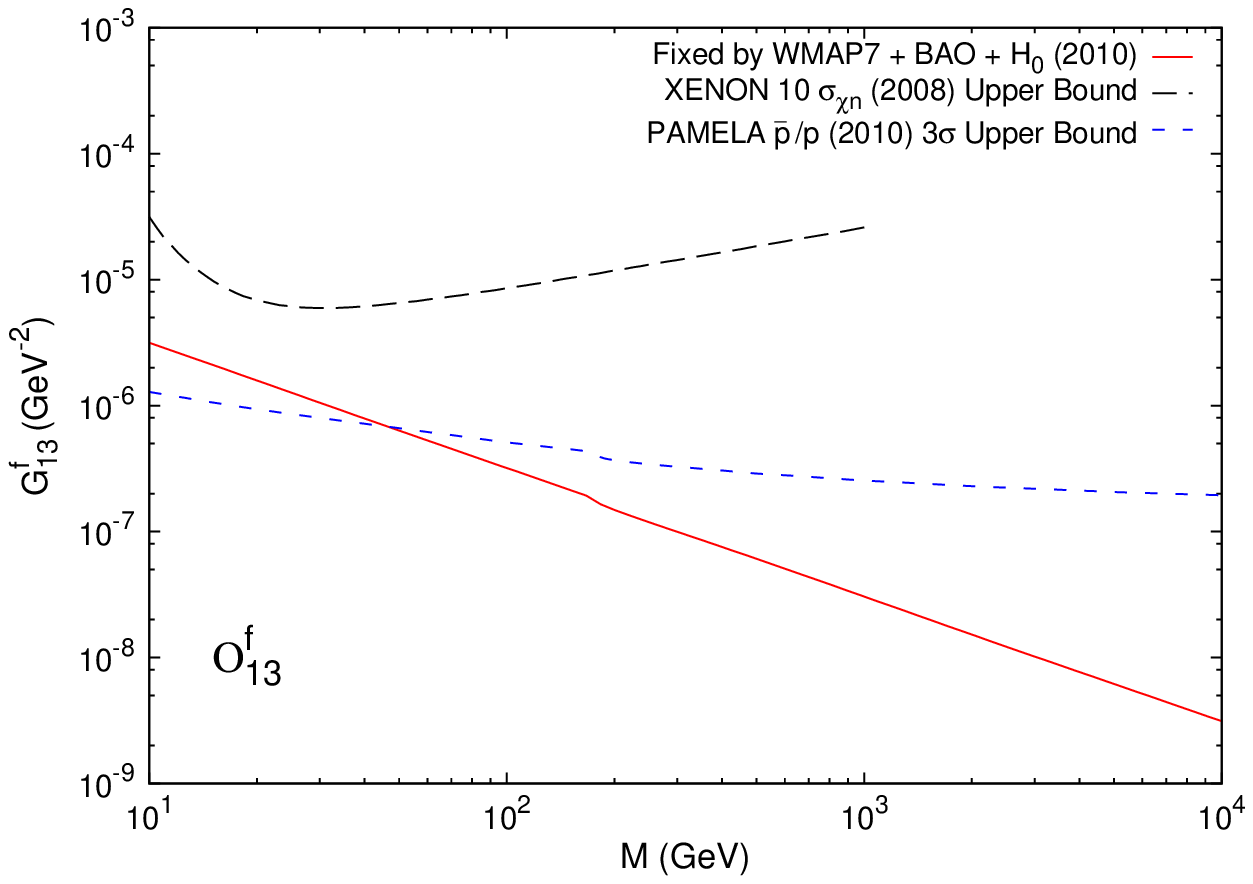}
\hspace{0.008\textwidth}%
\includegraphics[width=0.33\textwidth]{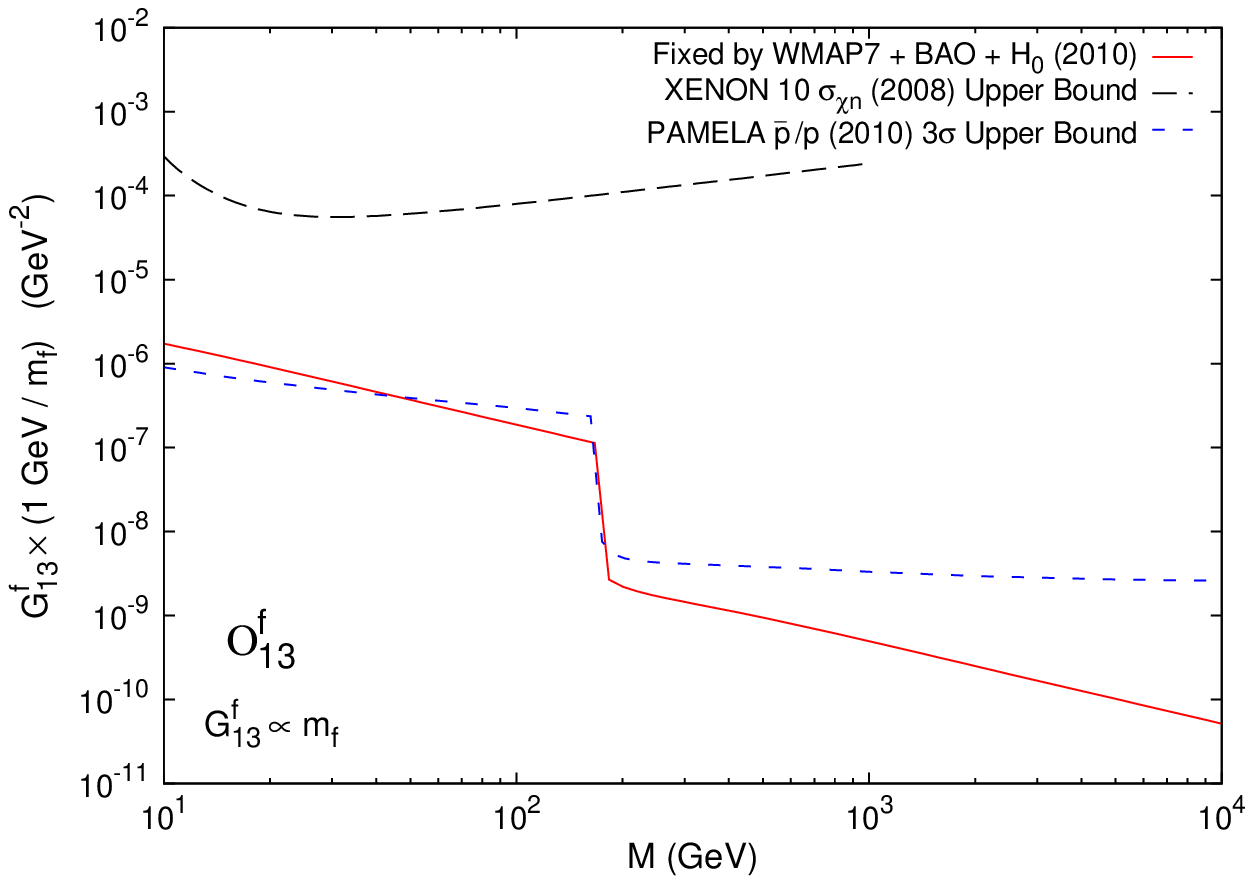}
\caption{Combined constraints on the couplings $G_{1,5,9,12,13}$ are
obtained as a function of $M$ from the observed relic density,
direct detection experiments of XENON10 (SD $\Psi$-neutron
scattering) and XENON100 (SI $\Psi$-proton scattering), and the
observed PAMELA $\bar p/ p$ flux ratio. Left panels for scenario I
and right ones for scenario II.%
\label{fig-combined}}
\end{figure}

In this section, we present the combined constraints from the
observed relic density, direct and indirect detection data discussed
in previous subsections. Since the direct detection is only
sensitive to the operators $\calO_{1,5,9,12,13}$, we show our
results for these types of interactions. In Fig \ref{fig-combined},
we show the combined constraints on their couplings
$G_{1,5,9,12,13}^{f}$ as a function of mass $M$ in both scenarios I
(left panels) and II (right ones). For the SI types of interactions
$\calO_{5,9}$ the direct detection generally imposes a stronger
constraint than the indirect detection except for $\calO_9$ in
scenario II. For the SD types of interactions the direct detection
is advantageous in some cases while the indirect detection is better
in others.

When the upper bound curves from direct detection (SI and SD) and
PAMELA $\bar p/p$ data are located below the relic-density allowed
curves, they can provide more stringent constraints and be used to
exclude some regions of parameters for various types of
interactions. In the scenario I of universal couplings, the PAMELA
data excludes  respectively the operators $\calO_{12,13}$ in the
mass ranges of $(10,17)~\GeV$ and $(10,47)~\GeV$ for any couplings.
While the SI direct detection completely excludes the operators
$\calO_{5,9}$, the SD one dominates for the operator $\calO_1$ and
excludes it in the mass range $(12,37)~\GeV$. The situation for
scenario II is similar but less stronger. For instance, $\calO_1$ is
not sensitive to either the SD direct detection or the PAMELA data
so that any mass would be allowed, while only a smaller mass range
$(10,49)~\GeV$ for the operator $\calO_9$ is excluded by PAMELA.

\section{Conclusion}

We have considered the option that a spin 3/2 particle acts as dark
matter, and investigated the constraints on it imposed by the
current observations and experiments. We worked in the approach of
effective field theory and wrote down all possible 4-fermion
effective interactions that involve a pair of spin-3/2 DM fields and
a pair of ordinary fermion fields. Assuming one interaction at a
time is responsible for the dark matter, we studied its implications
on the relic density, the antiproton to proton flux ratio $\bar p/p$
in cosmic rays, and the elastic dark matter scattering off nuclei in
direct detection. While the relic density and the flux ratio $\bar
p/p$ are virtually sensitive to all interactions at different levels
and for different scenarios of couplings, only a few are relevant to
the direct detection experiments. These observational and
experimental results can be employed in a complementary manner.
Using the observed relic density one can predict the relation
between the effective coupling and the DM mass for a given
interaction. When this relation curve lies above the upper bounds
set by the direct or indirect detection, we can exclude some of the
parameter regions for the DM particle. For example, the SI XENON100
data excludes the whole mass range that we studied for the
interactions $\calO_{5,9}^f$ when the couplings are flavor
universal. Depending on the types of interactions and scenarios of
couplings, the SD direct detection data and measured flux ratio
$\bar p/p$ can also exclude portions of mass ranges. Further precise
measurements will help narrow down the survival windows thus far.

\vspace{0.5cm}
\noindent %
{\bf Acknowledgement}

One of us (RD) would like to thank Prof. Xiaojun Bi for the
invitation of visits to the Institute of High Energy Physics, CAS,
and the members of his group for helpful discussions and generous
help with manipulating and developing codes. This work was supported
in part by the grant NSFC-11025525 and by The Fundamental Research
Funds for the Central Universities No.65030021.
\\

{\em Notes added}. While this manuscript was being finished, a new
preprint appeared \cite{Yu:2011by} in which the spin-3/2 particle
was also studied as a DM candidate. After this work was submitted to
the arXiv, we were informed of a recent paper \cite{Kamenik:2011vy}
in which effective operators involving a pair of spin-3/2 fields
were studied together with other operators involving the standard
model Higgs and gauge fields. A spin-3/2 particle was also proposed
earlier \cite{Khlopov:2008ki} in an attempt to reconcile puzzling
results in direct detections, as a charged effective degree of
freedom that is bound with primordial helium to form the so-called
dark atoms. We thank the authors of those papers for their
electronic communications.

\vspace{0.5cm}
\noindent %
{\bf Appendix} Functions $A_i(r,R)$

The functions $A_i(r,R)$ with $r=m_f^2/s$ and $R=M^2/s$ appearing
in the annihilation cross section are obtained upon finishing the
phase space integration:
\begin{eqnarray*}
A_1&=&-\frac{13}{54}  - \frac{46\,r}{27} + \frac{1}{108\,R^2} +
\frac{2\,r}{27\,R^2} - \frac{r}{3\,R} +
  \frac{10\,R}{27} + \frac{200\,r\,R}{27}
\\
A_2&=&-\frac{1}{54}  - \frac{34\,r}{27} + \frac{1}{108\,R^2} +
\frac{2\,r}{27\,R^2} + \frac{1}{27\,R} -
  \frac{7\,r}{27\,R}
\\
A_3&=&-\frac{13}{54} + \frac{20\,r}{27} + \frac{1}{108\,R^2} -
\frac{r}{27\,R^2} + \frac{r}{3\,R} +
  \frac{10\,R}{27} - \frac{160\,r\,R}{27}
\\
A_4&=&-\frac{1}{54} + \frac{32\,r}{27} + \frac{1}{108\,R^2} -
\frac{r}{27\,R^2} + \frac{1}{27\,R} +
  \frac{11\,r}{27\,R}
\\
A_5&=&\frac{7}{6} - \frac{14\,r}{3} + \frac{1}{36\,R^2} -
\frac{r}{9\,R^2} - \frac{5}{18\,R} + \frac{10\,r}{9\,R} -
  2\,R + 8\,r\,R
\\
A_6&=&\frac{5}{18} - \frac{10\,r}{9} + \frac{1}{36\,R^2} -
\frac{r}{9\,R^2} - \frac{1}{18\,R} + \frac{2\,r}{9\,R}
\\
A_7&=&\frac{7}{6} + \frac{1}{36\,R^2} - \frac{5}{18\,R} - 2\,R
\\
A_8&=&\frac{5}{18} + \frac{1}{36\,R^2} - \frac{1}{18\,R}
\\
A_9&=&-\frac{2}{27} - \frac{4\,r}{27} + \frac{1}{27\,R^2} +
\frac{2\,r}{27\,R^2} - \frac{2}{27\,R} -
  \frac{4\,r}{27\,R} + \frac{4\,R}{3} + \frac{8\,r\,R}{3}
\\
A_{10}&=&\frac{26}{27} + \frac{52\,r}{27} + \frac{1}{27\,R^2} +
\frac{2\,r}{27\,R^2} - \frac{8}{27\,R} -
  \frac{16\,r}{27\,R} - \frac{40\,R}{27} - \frac{80\,r\,R}{27}
\\
A_{11}&=&-\frac{2}{27}  + \frac{8\,r}{27} + \frac{1}{27\,R^2} -
\frac{4\,r}{27\,R^2} - \frac{2}{27\,R} +
  \frac{8\,r}{27\,R} + \frac{4\,R}{3} - \frac{16\,r\,R}{3}
\\
A_{12}&=&\frac{26}{27} - \frac{128\,r}{27} + \frac{1}{27\,R^2} -
\frac{4\,r}{27\,R^2} - \frac{8}{27\,R} +
  \frac{44\,r}{27\,R} - \frac{40\,R}{27} + \frac{280\,r\,R}{27}
\\
A_{13}&=&\frac{4}{27} - \frac{184\,r}{27} + \frac{2}{27\,R^2} +
\frac{4\,r}{27\,R^2} - \frac{4}{27\,R} +
  \frac{40\,r}{27\,R} + \frac{40\,R}{27} + \frac{800\,r\,R}{27}
\\
A_{14}&=&\frac{4}{27} + \frac{200\,r}{27} + \frac{2}{27\,R^2} +
\frac{4\,r}{27\,R^2} - \frac{4}{27\,R} -
  \frac{56\,r}{27\,R} + \frac{40\,R}{27} - \frac{640\,r\,R}{27}
\end{eqnarray*}

\noindent %

\end{document}